\documentclass[conference]{IEEEtran}
\IEEEoverridecommandlockouts
\usepackage{cite}
\usepackage{amsmath,amssymb,amsfonts}
\usepackage{graphicx}
\usepackage{textcomp}
\usepackage{xcolor}
\usepackage{algorithm}
\usepackage{algpseudocode}
\usepackage{booktabs}
\usepackage{multirow}
\usepackage{subfigure}
\usepackage{caption}
\usepackage{pifont}
\newcommand{\xmark}{\ding{55}}

\title{SecureLearn - An Attack-agnostic Defense for Multiclass Machine Learning Against Data Poisoning Attacks \\
\thanks{This work is submitted to IEEE Transactions on Information Forensics and Security.}}
\date{March 2025}
\makeatletter
\newcommand{\linebreakand}{%
  \end{@IEEEauthorhalign}
  \hfill\mbox{}\par
  \mbox{}\hfill\begin{@IEEEauthorhalign}
}
\makeatother
\begin{document}
\author{\IEEEauthorblockN{Anum~Paracha, Junaid~Arshad, Mohamed~Ben~Farah, Khalid~Ismail} 
\IEEEauthorblockA{\small College of Computing, Birmingham City University, Birmingham, United Kingdom}}

\maketitle

\begin{abstract} Data poisoning attacks are a potential threat to machine learning (ML) models, aiming to disrupt their learning processes by manipulating the training datasets. Existing defenses are mostly designed to mitigate specific poisoning attacks or are aligned with particular ML algorithms. Furthermore, most defenses are developed to mitigate poisoning attacks in deep neural networks or binary classifiers. However, traditional multiclass classifiers need attention to be secure from data poisoning attacks, as these models are significant in developing multi-modal applications, particularly with limited resources and feature-structured datasets. Therefore, this paper proposes SecureLearn, a two-layer attack-agnostic defense to defend multiclass models from poisoning attacks. It comprises two components of data sanitization and a new \textbf{f}eature-\textbf{or}iented adversarial \textbf{t}raining (FORT). To ascertain the effectiveness of SecureLearn, we proposed a 3D evaluation matrix with three orthogonal dimensions: data poisoning attack, data sanitization and adversarial training. Benchmarking SecureLearn in a 3D matrix, a detailed analysis is conducted at different poisoning levels (10\%-20\%), particularly analysing accuracy, recall, F1-score, detection and correction rates, and false discovery rate. The experimentation is conducted for four ML algorithms, namely Random Forest (RF), Decision Tree (DT), Gaussian Naive Bayes (GNB) and Multilayer Perceptron (MLP), trained with three public datasets: IRIS, MNIST and USPS, against three poisoning attacks and compared with two existing mitigation techniques. Our results highlight that SecureLearn is effective against the provided attacks in all given models. SecureLearn has strengthened resilience and adversarial robustness of traditional multiclass models and neural networks, confirming its generalization beyond algorithm-specific defenses. It consistently maintained accuracy above 90\%, recall and F1-score above 75\%, and reduced the false discovery rate to 0.06 across all evaluated models. In the context of neural networks, SecureLearn achieved at least 97\% recall and F1-score against all selected poisoning attacks. The adversarial robustness of models, trained with SecureLearn, improved with an average accuracy trade-off of only 3\%. 

\end{abstract}
\section{Introduction} In recent years, machine learning(ML) has been facilitating outstanding performance in prediction and decision-making tasks. For example, in a recommender system \cite{zhang2019deep}, biometric recognition \cite{jhong2020automated}, and security-sensitive applications such as skin cancer detection \cite{ghosh2024study}, medical imaging \cite{ma2021understanding} and autonomous vehicles \cite{chen2023edge}. ML models rely on training datasets to develop their decision-making mechanisms by identifying the underlying patterns in the given data and making predictions independently without additional information. \\
Despite their outstanding performance, recent studies show that these ML models are susceptible to various adversarial attacks, typically classified as data poisoning attack \cite{huang2020metapoison} which perturbs the training dataset, evasion attack \cite{QueryFree_Evasion_Attacks_GAN} which adds manipulations in test data, inversion attack \cite{Model_Inversion_Attacks_Graph_Neural_Networks} which tends to steal the confidential information of the model and inference attack \cite{Membership_Inference_Knowledge_Distillation} which tends to identify training dataset. 
Of these, we focus on data poisoning attacks in multiclass models, which pose serious security threats to ML. For example, outlier-oriented poisoning attack (OOP) \cite{OOPAttack} manipulates the feature space of the model by perturbing outliers, subpopulation attack(SubP) \cite{subpopulationattack} injects poisoned clusters into the dataset and exploits data sanitization techniques: TRIM \cite{Manipulating_ML} and SEVER \cite{diakonikolas2019sever}. Similarly, label-flipping attack \cite{xiao2012adversarial, shahid2022label} is a common data poisoning attack that can be extended as random label poisoning attack (RLPA) in multiclass models. 
Other successful data poisoning attacks are \cite{huang2020metapoison, baker2024poison, chen2021deeppoison}. \\
Recently, various defenses have been proposed to mitigate data poisoning attacks \cite{tao2021better, barreno2008open, paudice2019label}. However, these solutions are mostly attack-specific or system-specific, defined to mitigate specific data poisoning attacks or are adaptable to particular algorithms. For example, Hossain et al. \cite{hossain2024advancing} proposed a solution to detect backdoor attacks limited to deep neural networks. Baker et al. \cite{baker2024poison} developed a method to particularly secure recommender systems from data poisoning attacks, which does not defend other systems. Peri et al. \cite{peri2020deep} removed clean-label poison by detecting falsified data points with k-neighbors; however it is only effective against feature collision and convex polytope attacks. Adversarial training \cite{ho20224364}, a prominent adversarial defense, is only adaptable in deep learning (DL) as it follows gradient learning. Moreover, various attacks, such as \cite{koh2021stronger, Venkatesan2021874, ge2023advancing}, have successfully breached defenses against data poisoning attacks with evolving attack vectors. Currently, few solutions are proposed that offer attack-agnostic defense, and these solutions are mostly designed for DL models. \\
Given the above-mentioned limitations, we propose SecureLearn, a two-layer attack-agnostic approach to defend against data poisoning, irrespective of particular attack vectors. 
SecureLearn offers an enhanced data sanitization that combines the fundamental principles of nearest neighbor voting strategy to correct data labels, followed by calculating the statistical deviations of each data point to detect and correct anomalies. Furthermore, SecureLearn introduced a new approach of \textbf{f}eature-\textbf{or}iented adversarial \textbf{t}raining (FORT) influenced by a common characteristic of feature importance score of ML to identify important data points to generate adversarial examples for training. \\
To thoroughly assess SecureLearn, we propose a 3D evaluation matrix following three dimensions: data poisoning attacks, data sanitization and adversarial training. The experiments are conducted on four machine learning algorithms: Random Forest (RF), Decision Tree (DT), Gaussian Naive Bayes (GNB), and Multi-layer Perceptron (MLP). Selecting these algorithms allows us to cover most types of classification mechanisms in machine learning. We selected three distinct data poisoning attacks: OOP, SubP and RLPA attacks and set the poisoning levels between 10\% and 20\% at a scale of 5 to study the effectiveness of SecureLearn in different adversarial settings. We also compare it with two data sanitization defenses given in \cite{chan2018data, paudice2019label}, highlighting enhanced performance and generalization of SecureLearn over others. The contributions of this paper are given as follows:
\begin{itemize}
\item To the best of our knowledge, SecureLearn is the first attack-agnostic defense in multiclass classifiers defending against data poisoning attacks. SecureLearn provides defense without requiring prior knowledge of attacks, targeted models and additional data.
\item We have proposed a new adversarial training mechanism named Feature-Oriented Adversarial Training (FORT) as a component of SecureLearn, enhancing the adversarial robustness of traditional multiclass ML, including neural networks. Our results show that the adversarial robustness improved with a minimal trade-off between accuracy and robustness, i.e., the accuracy is decreased $<3\%$, while enhancing the adversarial robustness.  
\item We have proposed a new 3D evaluation matrix to comprehensively evaluate SecureLearn against three data poisoning attacks and compare it with two existing defenses. The evaluation is set up for four types of ML models trained with three distinct datasets. The results highlight that SecureLearn has outperformed other mitigations and is effective against all selected attacks for all models, consistently maintaining accuracy to a minimum 90\% and recall and F1-score to 75\%.
\end{itemize}
            

\section{Related Work} \subsection{Existing Multiclass Poisoning Attacks} The existing literature highlights various data poisoning attacks that affect the confidentiality, integrity, and availability of multiclass models. Such as Alarab et al. \cite{alarab2023uncertainty} experimentally showed an increase in model variance and prediction uncertainty with a manipulated dataset. They also highlight the limitations of the Monte-Carlo method in detecting poisoned data points near classification boundaries. MetaPoison \cite{huang2020metapoison} manipulates the training dataset to fool neural networks. This attack craft poisoned images by solving bilevel optimization with the Carlini and Wagner attack \cite{aditya2024adversarial} and achieved a 40-90\% success to poison all selected models with a 1\% poison budget. They also experimented the MetaPoison on Google Cloud AutoML API and achieved $>15\%$ success with a minimum of 0.5\% dataset poisoning. Zhao et al. \cite{zhao2022clpa} proposed a class-oriented targeted attack to manipulate individual classes in DL models, whereas Lu et al. \cite{lu2023exploring} introduced model poisoning reachability to quantify the limits of targeted poisoning. Munoz-Gunzalez et al. \cite{munoz2017towards} extended gradient optimization poisoning in multiclass DL models. \\
Alongside poisoning modern ML models, certain attacks are introduced to manipulate traditional multiclass models. OOP attack \cite{OOPAttack} manipulated the feature space of multiclass models by exploiting outliers in the dataset and experimented against six models. Biggio et al. \cite{10.5555/3042573.3042761} introduced an adversarial label flipping attack to poison class labels indiscriminately, which can be extended to poison multiclass datasets. Jagielski et al. \cite{10.1145/3460120.3485368} introduced a clean label poisoning that augments a cluster of poisoned points in the training dataset, challenging poison detection as it is difficult to identify a subset of poisoned data points with similar features. Pantelakis et al. \cite{pantelakis2023adversarial} experimented JSMA, FGSM and DeepFool attacks to evaluate performance disruption in multiclass IoT networks. 
\subsection{Limitations of Existing Solutions}
In contrast, various mitigation techniques are proposed in the literature to secure ML from data poisoning attacks. Such as Neehar et al. \cite{peri2020deep} developed a deep k-NN to remove clean label poison by detecting falsified data points with k-neighbors. Deep k-NN defense is experimented against feature collision and convex polytope in deep neural networks. Paudice et al. \cite{paudice2019label} used the k-NN algorithm to mitigate label poisoning in binary SVM. Carnerero-Cano et al. \cite{carnerero2023hyperparameter} computed limitations of hyperparameters to resist data poisoning impact on DNN models. Barreno et al. \cite{barreno2008open} have given the concept of reject on negative impact to remove affected data points, which is extended in \cite{chan2018data} to filter poisoned data from the given dataset. However, most mitigations are implemented to secure either DL models or are applicable to binary ML models. Limited solutions are provided to secure traditional multiclass models, such as one-versus-one SVM, multiclass GNB, RF or DT algorithms. We also need a mitigation mechanism that is attack-agnostic and can be adaptable to secure ML from evolving data poisoning attacks, i.e., effective against most types of existing and novel data poisoning attacks and can be implemented with various datasets and algorithms in both binary and multiclass settings. \\
Adversarial training is a prominent defense to improve the adversarial robustness of DL models. Such as \cite{ho20224364,tao2021better,shafahi2020universal} have proposed adversarial training methods and implemented in neural networks and DL models as adversarial training is designed following iterative gradient learning, which does not apply to traditional models hence makes adversarial training ineffective in securing traditional ML models. \\  
Conclusively, there are some attack-agnostic solutions proposed in the literature that are mentioned to be adaptable to various data poisoning attacks, while mostly focused on securing DL models. To secure traditional ML, few solutions are proposed; however, these mitigations are experimented to improve the robustness of binary models; however, limited attention is given to traditional models developed in multiclass settings. \\
SecureLearn is an attack-agnostic solution which is designed to be adaptable to traditional ML and neural networks in multiclass classification settings. It is effective against various aforementioned attacks, providing promising results in various real-world applications. 
SecureLearn is proposed as a two-layer solution with improved data sanitization and a feature-oriented adversarial training to strengthen model robustness. A brief comparison of various existing solutions with SecureLearn is provided in Table \ref{tab:existing literature}, highlighting that existing solutions have either proposed data sanitization or adversarial training, where data sanitization solutions are experimented on binary ML models and adversarial training is experimented with only DL models.
\begin{table*}[h]
\scriptsize 
    \centering
    \caption{Summary of existing similar defenses against data poisoning attacks proposed in various settings}
    \label{tab:existing literature}
    \begin{tabular}{c c c c c}
\toprule
Research paper & Data Sanitization & Adv. Training & ML model & Model Settings \\ 
\midrule
De-Pois \cite{chen2021pois} & \checkmark & \xmark & GAN, CNN and LASSO & Binary and Multiclass DNN \\
A. Paudice et al. \cite{paudice2019label} & \checkmark & \xmark & Stochastic Gradient Descent & Binary ML \\
P. PK Chan et al. \cite{chan2018data} & \checkmark & \xmark & SVM & Binary ML \\
M. Barreno et al. \cite{barreno2008open} & \checkmark & \xmark & SVM & Binary ML \\
A. Shafahi et al. \cite{shafahi2020universal} & \xmark & \checkmark & ResNet and InceptionV1 & Multiclass DNN \\
L. Tao et al. \cite{tao2021better} & \xmark & \checkmark & VGG-16, VGG-19, ResNet-18, ResNet-50 and DenseNet-121 & Multiclass DL \\
\textbf{SecureLearn} & \checkmark & \checkmark & DT, RF, GNB, NN & Multiclass ML \\
\bottomrule 
    \end{tabular}
\end{table*}    
\section{Threat Model}
\subsection{Attacker's Goal} We defined two attacker goals to assess the effectiveness of selected mitigation solutions. The first goal is to disrupt the model's availability and reduce its overall performance by employing the OOP attack \cite{OOPAttack} and label flipping attack \cite{shahid2022label}. The second goal is to harm the model's integrity by augmenting clustered poisoned data points employing the subpopulation attack to disrupt targeted class predictions \cite{subpopulationattack}. \\
Consider the poisoning of supervised classification models, e.g. RF or MLP, given the dataset $D_o = \{(x_{i}, l_{i})\}_{i=1}^n$ with data points $x$ and labels $l$, the attacker can manipulate the labels $l'$ or the features $x'$ of the dataset or augment poisoned data points($x'$, $l'$) into the dataset to prevent the trained victim model from attaining the intended performance. 
\subsection{Attacker's Knowledge} In this threat model, the attacker possesses limited knowledge of the targeted model $M$ and dataset $D_{o}$. Under these constraints, all selected data poisoning attacks are formulated as gray-box attacks. In this scenario, the attacker has a partial understanding of the dataset and model: the dataset and algorithm names are known, but the dataset distribution, model settings, and parameters remain unknown. Additionally, the attacker has no knowledge and access to the target system.  
\subsection{Attacker's Capability} We have leveraged the attacker's capability to poison the training datasets in different ways. The attacker can modify labels or features of the dataset and introduce poisoned data points into the dataset. However, this capability is limited to injecting a maximum 20\% poisoning level as the upper bound limit and a minimum 10\% poisoning as the lower bound limit. These limits are defined as the most effective poisoning limits \cite{DataPoisonBehaviour, OOPAttack}, highlighting $10\% \leq \Delta L \leq 20\%$ are complacent poisoning levels, whereas $\Delta L <5\%$ has a negligible impact and $\Delta L>20\%$ is detectable. 
\subsection{Attack Strategy} In our attack settings, three data poisoning attacks of varying attack vectors, i.e., OOP, SubP and RLPA attacks are considered. Following these attacks in multiclass classifiers, the effectiveness of SecureLearn is evaluated, demonstrating that it is an attack-agnostic and promising solution capable of mitigating all the aforementioned attacks.
\section{SecureLearn Design} 
\begin{figure*}[h]
  \centering      
  \includegraphics[width=\linewidth]{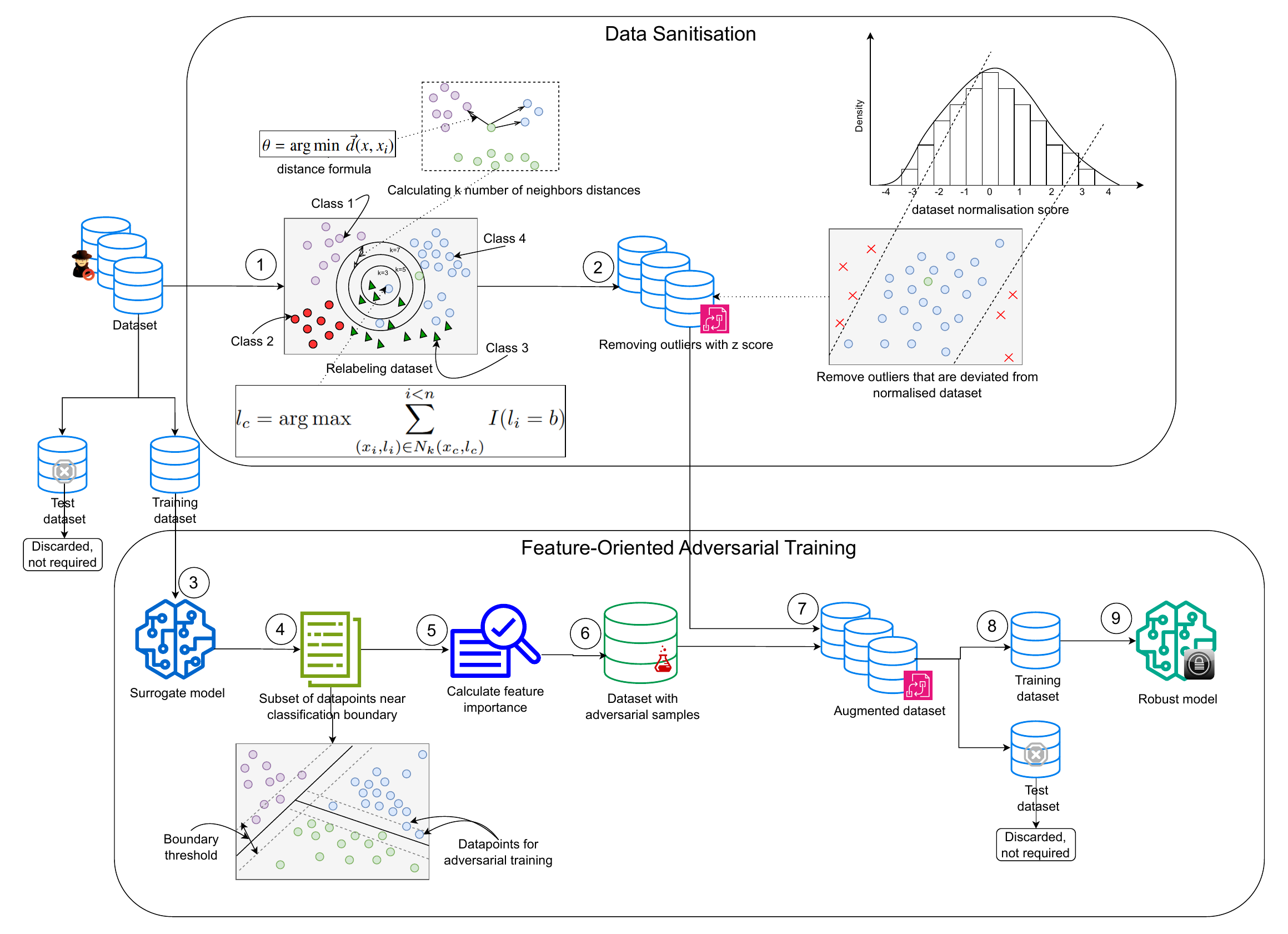}
  \caption{Architectural overview of SecureLearn illustrating a two-layer approach to secure the training pipeline of ML models irrespective of data poisoning attacks}

  \label{solution}
\end{figure*}

We formulate the problem of poisoning the training dataset, given as follows: $D_c$ is the clean dataset, $D_c'$ is the poisoned substitute in the dataset formulated as $D_o = D_c \cup D_c'$. As no ground truth is provided, SecureLearn aims to sanitize $D_o$ to correct data points and align features. 
SecureLearn relies on the general observation that the poisoned dataset tricks the model training to classify differently from the clean dataset, resulting in performance degradation. Therefore, SecureLearn identifies anomalies and misalignments in the features of the data points and their labels. Furthermore, SecureLearn improves the resilience of the model with adversarial training. To achieve this aim, SecureLearn comprises the following two components: data sanitization and feature-oriented adversarial training. The complete process to improve the resilience of the ML model with SecureLearn is illustrated in Fig. \ref{solution}. The algorithm of SecureLearn is provided in Alg. \ref{securelearn algorithm}. 
\begin{algorithm}
\caption{SecureLearn Mitigation Mechanism }
\label{securelearn algorithm}
\begin{algorithmic}
\State \textbf{Input}: Training samples \textit{X}, perturbation limit $\varepsilon$, feature importance scores: \textit{$F$} 
\State \textbf{Initialize}: b=0.001, c=0.01, nearest neighbors (k)=7
\For{$x_{i} \in X$}
\State $d = min(k, dist(x_{i}, x))$
\State $l_{i} = avg(x_{i}, d)$
\State $D_{san} \gets (x_{i}, l_{i})$
\EndFor
\For{$x_{i} \in D_{san}$}
\State $Compute \: \delta_{i} \: following \: Eq. 5.5$
\If{$\delta < |g|$}
\State $D_{san} \gets (x_{i}, l_{i})$
\EndIf
\EndFor
    \If{$M == M_{GNB}\: or \: M_{MLP}$}
    \State $F \gets \arg max \: Probability(D_{san})$
    \EndIf
    \If $(M == M_{RF} \: or \: M_{DT})$:

    \State $F \gets \sum_{i=1}^{L} f_{i}(1 - f_{i})$
    \EndIf
    \For{($x_{i} \in D_{san}$) and ($f_{i} \in F$)}
    \State $D_{adv} \gets \mathbb{E}_{(x, y) \sim D_{o}} [\mathcal{L} (M, (x_{i}+ (c * sign((f_{i}*x_{i}) + b)))$
    \EndFor
\end{algorithmic}
\end{algorithm}
\subsection{Data Sanitization} 
Our data sanitization module comprises two parts: relabeling the dataset $D_{o}$ and removing anomalies. Our relabeling mechanism is defined as:
\begin{equation}
\begin{aligned}
D_{san} = \{(x, l)|x \in D_{o}\} \\
and \: l = \begin{cases}
l_{i} \: if \: C(x_{i}, l_{i}) < \gamma \\
l \: if \: C(x_{i}, l_{i}) \geq \gamma
\end{cases}
\end{aligned}
\label{Eq:LabelSanitisation}
\end{equation}
where $D_{san}$ is the sanitized dataset, $C(x, l)$ is the confidence of neighboring data points, $l_{i}$ is the existing label of the data point $x$, and $l$ is the new label confident label from the nearest data points. The confidence limit is defined as $\gamma \geq 40\%$ neighboring votes, following an incremental majority voting approach \cite{abdulboriy2024incremental}. The calculation of the label of each data point, given in Eq. \ref{Eq:LabelSanitisation}, follows the confidence score $C(x,l)$ of neighboring data points, calculated with Eq. \ref{Eq:confidence_score}. 
\begin{equation}
C(x, l) = \arg \max \frac{1}{k} \sum_{(x_j, l_j)\in \theta}^{j<n}{I}(l_{j} = l_c)\\
\label{Eq:confidence_score}
\end{equation}
where $l_{c}$ is the original class label, $k$ is the no. of nearest neighbors set to seven following the kTree method given by \cite{zhang2017efficient}, $x$ is the data point with label $l$ and $\theta$ is the function of measure of distance given in Eq. \ref{Eq:distancevector}.
\begin{equation}
\theta = \min \vec{d}(x_{i}, x)\\
\label{Eq:distancevector}
\end{equation}
The next part of data sanitization is to remove outliers from the dataset. The anomalous data points are removed from the dataset, where the deviation ($\delta$) of the given data point exceeds the limits of the normalized dataset distribution, following Eq. \ref{Eq:removeoutliers}. The $\delta$ is calculated with Eq. \ref{Eq:zscore} where $\mu$ is the mean of the dataset and the deviation limit $|g|=3$ \cite{abdi2007z}. 
\begin{equation}
D_{san} = \{x_{i} \in D_{o} || \delta \leq |k|\}
\label{Eq:removeoutliers}
\end{equation} 
\begin{equation}
\delta = \frac{x_{i} - \frac{1}{n} \sum_{i=1}^n{x_{i}}}{\sqrt{\frac{1}{n} \sum_{i=1}^n{(x_{i} - \mu)^2}}}
\label{Eq:zscore}
\end{equation}          
\subsection{Feature-Oriented Adversarial Training (FORT)} After obtaining the sanitized dataset, SecureLearn aims to improve the adversarial robustness of the model with feature-oriented adversarial training. In the literature, it is noticed that the existing adversarial training mechanism is unable to improve the resilience of traditional ML models \cite{paracha2024exploring} because existing approach follows the gradient-oriented training which is ineffective for traditional models, therefore SecureLearn introduced a new method to train models, where adversarial data $D_{adv}$ is generated by augmenting data points with high feature importance score and lie near the decision boundary. This is done by solving Eq. \ref{Eq:dataaugmentation}, followed by generating the perturbation in Eq. \ref{Eq:dataperturbation}.
\begin{equation}
D_{adv} \gets \mathbb{E}_{(x, y) \sim D_{o}} [\mathcal{L} (M, ((x+\varepsilon), l)
\label{Eq:dataaugmentation}
\end{equation}
where $M$ is the training model, $\mathcal{L}$ is the training loss and  $\varepsilon$ is the perturbation given in Eq. \ref{Eq:dataperturbation}.
\begin{equation}
 \varepsilon = c * sign((f_{i}*x_{i}) + b)
\label{Eq:dataperturbation}
\end{equation}
where, in Eq. \ref{Eq:dataperturbation}, $f_{i}$ is the feature importance score of the model $\textit{M}$, $c=0.01$ is the perturbation constant, following the average perturbation value given in \cite{liu2021model}. $x_{i}$ is the data point, and $b=0.001$ is the non-zero coefficient. Combining output of Eq. \ref{Eq:LabelSanitisation} and Eq. \ref{Eq:dataaugmentation}, the sanitized dataset $D_{s}$ is given in Eq. \ref{Eq:trainingdataset}: 
\begin{equation}
D_{s} = D_{san} + D_{adv}
\label{Eq:trainingdataset}
\end{equation}
Intuitively, the model is trained to mitigate the data poisoning effects and improve the overall performance. 
Unlike traditional adversarial training based on gradient optimization, FORT adds slight perturbations to the data points that are close to the decision boundaries of the model to widen these boundaries, making them robust to poisoning. This way, SecureLearn improves the security and robustness of ML models against data poisoning attacks. To assess the effectiveness of SecureLearn, the evaluation matrix is described in Section \ref{evaluation matrix}.
\section{Experimentation And Ablation Study}
\subsection{Experimental Setup} We build our test environment and implement all attacks and defense techniques in Python using scikit-learn packages and NumPy API. All experiments are run on a 56-core Intel(R) Xeon(R) Gold 6258R CPU @ 2.70 GHz machine. In the experiment, we randomly split the dataset into 75\% for training and 25\% for testing after implementing the defense.   
\subsection{Datasets} We implemented all the attacks with three datasets of IRIS, MNIST and USPS. They have been widely used in studies of data poisoning attacks \cite{drews2020proving, wang2021robust, OOPAttack} and defenses \cite{xu2021detecting, jia2021intrinsic}. For each dataset, we implement each attack with three poisoning levels $\Delta L= (10,15,20)\%$. 
Selecting these datasets allows us to analyze the effectiveness of SecureLearn for differently structured datasets. Datasets structure is provided in Table \ref{tab:dataset} and their features association and correlation is given in Table \ref{tab:stat_corr}. 
\begin{table}[h]
\small
    \centering
    \caption{Dataset description}
    \label{tab:dataset}
    \begin{tabular}{c c c c}
\toprule
Dataset & No. of classes & No. of features & No. of instances \\
\midrule
IRIS & 3 & 4 & 150 \\
MNIST & 10 & 784 & 70,000 \\
USPS & 10 & 256 & 9298 \\
\bottomrule 
    \end{tabular}
\end{table}
\begin{table}[h]
\small
    \centering
    \caption{Features correlation in dataset}
    \label{tab:stat_corr}
    \begin{tabular}{c c c}
\toprule
Dataset & Spearman correlation & p-value \\
\midrule
IRIS & 0.1238 & 0.0791 \\
MNIST & 0.009282 & 0.0141 \\
USPS & -0.008742 & 0.2397 \\
\bottomrule 
    \end{tabular}
\end{table}
 
\subsection{3D Evaluation Matrix} \label{evaluation matrix} 
\begin{figure}[h]
  \centering      
  \includegraphics[width=\linewidth]{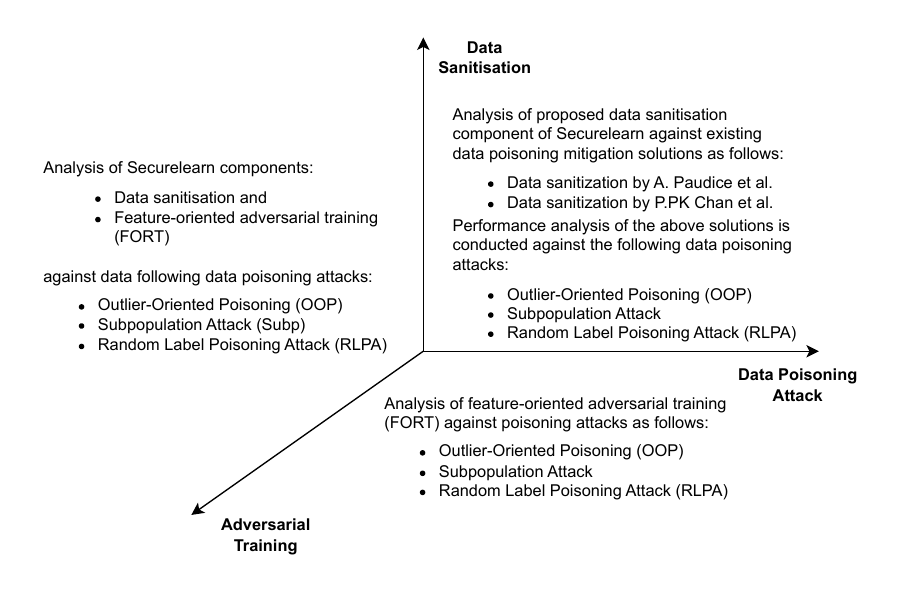}
  \caption{3D evaluation matrix to evaluate SecureLearn from three different aspects}

  \label{evaluation_matrix}
\end{figure}

We evaluate the SecureLearn in three dimensions, compare it with two typical defenses against first three data poisoning attacks as given in Table \ref{tab:existing literature}. The 3D evaluation matrix is illustrated in Fig. \ref{evaluation_matrix}. Its dimensions are explained as follows. 
\subsubsection{Dimensional Space 1} In dimensional space 1 (DS1), lies between data sanitization and data poisoning attack, we analyzed SecureLearn by experimenting with it against three data poisoning attacks and by comparing it with two existing similar defenses. The DS1 evaluates the strength of SecureLearn as an attack-agnostic defense to data poisoning attacks, followed by highlighting the profound performance of SecureLearn compared to other solutions. 
\subsubsection{Dimensional Space 2} In dimensional space 2 (DS2), lies between the dimensions of data poisoning attacks and adversarial training, we assess the effectiveness of the proposed FORT training component of SecureLearn against selected data poisoning attacks and analyzing improvements in the adversarial robustness of the model. 
\subsubsection{Dimensional Space 3} In dimensional space 3 (DS3), which lies between the dimensions of adversarial training and data sanitization, we assess the overall effectiveness of SecureLearn in securing multiclass ML from data poisoning attacks. It analyzed the false discovery rate of the model at varying poisoning levels against selected data poisoning attacks.   
\subsection{Evaluation Metrics} To evaluate model performance in a 3D evaluation matrix, we adopted the standard performance metrics: Accuracy, Recall and F1-score. Furthermore, the detection rate (DR), correction rate (CR) and false discovery rate (FDR) are utilized for the detailed evaluation. The DR and CR prominently highlight the efficacy of SecureLearn in sanitizing poisoned data points and FDR highlights the strengthened robustness of the model against poisoned training. Accuracy is the measure of correct classifications, where the poisoned data points remain disjointed in the incorrect classes and do not affect the model's availability. Recall measures the correct predictions of positive classifications over all positive answers, defining high separability. F1-score quantifies the overall defense performance, where the decision boundaries are aligned. Let the classification function be given in Eq. \ref{Eq:classificationfunction}, the evaluation metrics can be found in Eq. \ref{Eq:accuracy}, \ref{Eq:recall}, \ref{Eq:f1_score}.
\begin{equation}
f(C(x_{t})) = \begin{cases}
    true & \text{if $x_{t} \in Class \; c$}\\
    false & \text{otherwise}
\end{cases}
\label{Eq:classificationfunction}
\end{equation}
where $f$ is the classification function, $x_{t}$ is the data point from the test dataset $D_{t}$ split from $D_{s}$, and $C(.)$ is the class predictor. After sanitizing dataset with SecureLearn, false positives(FP) is defined as $f_{tr}(C(x_{{t}_{i}})|l_{c}')$, where $l_{c}'$ is the wrong class label and false negative(FN) is defined as $f_{fs}(C(x_{{tr}_{i}})|l_{c})$ where data points are not sanitized correctly. Whereas, true positive is defined as $f_{tr}(C(x_{{t}_{i}}))$ and true negative is defined as $f_{fs}(C(x_{{tr}_{i}}))$.
\begin{equation}\label{Eq:accuracy}
Acc = \frac{\sum_{i=0}^n f_{fs}(C(x_{{t}_{i}}))\land \sum_{i=0}^n f_{tr}(C(x_{{t}_{i}}))}{(x_{t} \in D_{t})}
\end{equation}

\begin{equation}\label{Eq:recall}
\begin{aligned}
Rcl = \frac{\sum_{i=0}^n f_{tr}(C(x_{{t}_{i}}))}{\sum_{i=0}^n(f_{tr}(C(x_{{t}_{i}}))) \land \sum_{i=0}^n(f_{fs}(C(x_{{t}_{i}})))}\\\\
where\: f_{fs}(C(x_{{t}_{i}})) \in D_{t}
\end{aligned}
\end{equation}

\begin{equation}\label{Eq:f1_score}
\resizebox{\hsize}{!}{$
\begin{aligned}
F1\_scr = \frac{\sum_{i=0}^n f_{tr}(C(x_{{t}_{i}})) * Rcl}{2 * \{(\sum_{i=0}^nf_{tr}(C(x_{{t}_{i}})) \land \sum_{i=0}^nf_{tr}(C(x_{{t}_{i}}))) + Rcl\}}
\end{aligned}$}
\end{equation}
Let $x'$ be the poisoned data point in $D_{o}$, and detection of these points with SecureLearn is given in Eq. \ref{Eq:detection_rate}, and setting these points in the appropriate class is shown in Eq. \ref{Eq:correction_rate}. After corrections, we analyze the false discovery rate of the model with Eq. \ref{Eq:fdr}.

\begin{equation}\label{Eq:detection_rate}
\begin{aligned}
DR = \frac{\sum_{i=0}^{n} P(x'|l_{c})}{\sum_{i=0}^{n}{ P(x|l_{c}) \land P(x'|l_{c})}}
\end{aligned}
\end{equation}

\begin{equation}\label{Eq:correction_rate}
\begin{aligned}
CR = \frac{\sum_{i=0}^{n}{P(x' \to x|ll_{c})}}{\sum_{i=0}^{n} P(x|l_{c}) \land P(x'|l_{c})}
\end{aligned}
\end{equation}

\begin{equation}\label{Eq:fdr}
\begin{aligned}
FDR = \frac{\sum_{i=0}^{n} f_{tr}(C(x_{t_{i}})|l_{c}')}{\sum_{i=0}^{n} f_{tr}(C(x_{t_{i}}|l_{c}')) \land f_{tr}(C(x_{t_{i}}))}
\end{aligned}
\end{equation}

\subsection{Experimental Results And Analysis} We conducted the experimental evaluation of SecureLearn with the 3D evaluation matrix defined in Fig. \ref{evaluation_matrix}. Our objective is to analyze the effectiveness of SecureLearn and understand its efficacy compared to existing solutions. We specifically answer how SecureLearn is better in detecting and sanitizing various types of poisons under \textit{DS1}, given in Sections \ref{boundaries} and \ref{data_sanitisation}. Furthermore, we understand how FORT is effective in generalizing traditional ML models and neural networks under \textit{DS2}, given in Section \ref{fo_trn}. We also understand the relationship between the increasing poisoning rate and resilience provided by SecureLearn under \textit{DS3}, given in Section \ref{poisoning_rate}.     
\subsubsection{\textbf{Determining Detection And Correction Boundaries}} \label{boundaries} We begin our analysis by determining the detection and correction rates against each data poisoning attack given in Table \ref{tab:bounds}. We calculate the lower bound (LB) and upper bound (UB) of DR and CR for each dataset at three defined poisoning levels from Eq. \ref{Eq:detection_rate} and Eq. \ref{Eq:correction_rate}, respectively. Our findings indicate that SecureLearn has detected at least 50\% poison from trained models regardless of the poisoning attack and the dataset being used. We observe that the minimum CR is ~30\% for the RF model against the RLP attack, likely due to the unpredictable placement and impact of poisoned data points in untargeted attacks. However, the UB of DR and CR of SecureLearn reaches 100\% to mitigate selected attacks trained with the IRIS dataset for most algorithms. We observe that SecureLearn is highly effective in sanitizing the IRIS dataset followed by the USPS dataset, compared to MNIST dataset, across all poisoning levels. This shows an inverse relation between SecureLearn performance and the dataset size. SecureLearn is generalizable across different poisoning strategies and dataset structures, performing independent to the no. of classes in the dataset. 
\begin{table*}[h]
\small
    \centering
    \caption{Detection and correction boundaries of individual ML models after mitigating data poisoning attacks with SecureLearn}
    \label{tab:bounds}
    \begin{tabular}{c c c c c c c c c}
\toprule
\multirow{3}{*}{Algorithm} & \multirow{3}{*}{Dataset} & & & \multicolumn{3}{c}{Attack}\\
\cline{4-9}
& & & \multicolumn{2}{c }{OOP} & \multicolumn{2}{ c }{Subp} & \multicolumn{2}{ c}{RLP}\\
\cline{4-5} \cline{6-7} \cline{8-9}
& & & LB & UB & LB & UB & LB & UB \\
\midrule
\multirow{6}{*}{RF} & \multirow{2}{*}{IRIS} & DR & 86.6 & 100 & 86.6 & 100 & 76.6 & 100 \\
& & CR & 80 & 90.9 & 80 & 91 & 76.6 & 93.3 \\
& \multirow{2}{*}{MNIST} & DR & 56.3 & 65.5 & 56.3 & 66.3 & 52.4 & 66.3 \\
& & CR & 33.5 & 49.2 & 33.5 & 49.2 & 29.7 & 47.6 \\
& \multirow{2}{*}{USPS} & DR & 87.94 & 89.13 & 56.29 & 65.78 & 50.48 & 62.56 \\
& & CR & 44.47 & 49 & 38.42 & 44.54 & 35.22 & 43.24 \\\hline
\multirow{6}{*}{DT} & \multirow{2}{*}{IRIS} & DR & 83.3 & 93.3 & 83.1 & 92 & 93.3 & 95.4 \\
& & CR & 86.6 & 90.9 & 80 & 91 & 76.6 & 91 \\
& \multirow{2}{*}{MNIST} & DR & 49.6 & 66.7 & 49.8 & 66.7 & 46.4 & 64.1 \\
& & CR & 44.69 & 57.88 & 45.1 & 58 & 44.97 & 55.08 \\
& \multirow{2}{*}{USPS} & DR & 44.69 & 57.88 & 44.69 & 57.88 & 44.97 & 55.08 \\
& & CR & 15.98 & 36.93 & 15.98 & 37 & 18.1 & 34.51 \\\hline
\multirow{6}{*}{GNB} & \multirow{2}{*}{IRIS} & DR & 100 & 100 & 100 & 100 & 80 & 100 \\
& & CR & 93.3 & 100 & 93.3 & 100 & 66.6 & 93.3 \\
& \multirow{2}{*}{MNIST} & DR & 98.6 & 99.1 & 98.6 & 99.1 & 96 & 98.4 \\
& & CR & 94.9 & 95.9 & 94.9 & 95.9 & 92.4 & 95.3 \\
& \multirow{2}{*}{USPS} & DR & 99.24 & 99.71 & 99.24 & 99.71 & 97.09 & 99.49 \\
& & CR & 97.63 & 97.99 & 97.63 & 97.99 & 95.53 & 97.99 \\\hline
\multirow{6}{*}{NN} & \multirow{2}{*}{IRIS} & DR & 83.3 & 100 & 83.3 & 100 & 73.3 & 95.4 \\
& & CR & 76.6 & 95.4 & 70 & 95.4 & 66.6 & 86.6 \\
& \multirow{2}{*}{MNIST} & DR & 56.3 & 65.5 & 56.3 & 66.3 & 52.4 & 66.3 \\
& & CR & 59.33 & 49.2 & 33.5 & 49.2 & 29.7 & 47.6 \\
& \multirow{2}{*}{USPS} & DR & 71.16 & 85.36 & 70.79 & 84.7 & 64.28 & 82.5 \\
& & CR & 59.33 & 78.9 & 59.11 & 79.76 & 51.47 & 76.42 \\
\bottomrule 
    \end{tabular}
\end{table*}
\subsubsection{\textbf{SecureLearn vs Existing Defenses}} \label{data_sanitisation} 
\begin{figure*}[!t]
\centering
\subfigure
    {\includegraphics[width=3cm,height=3cm]{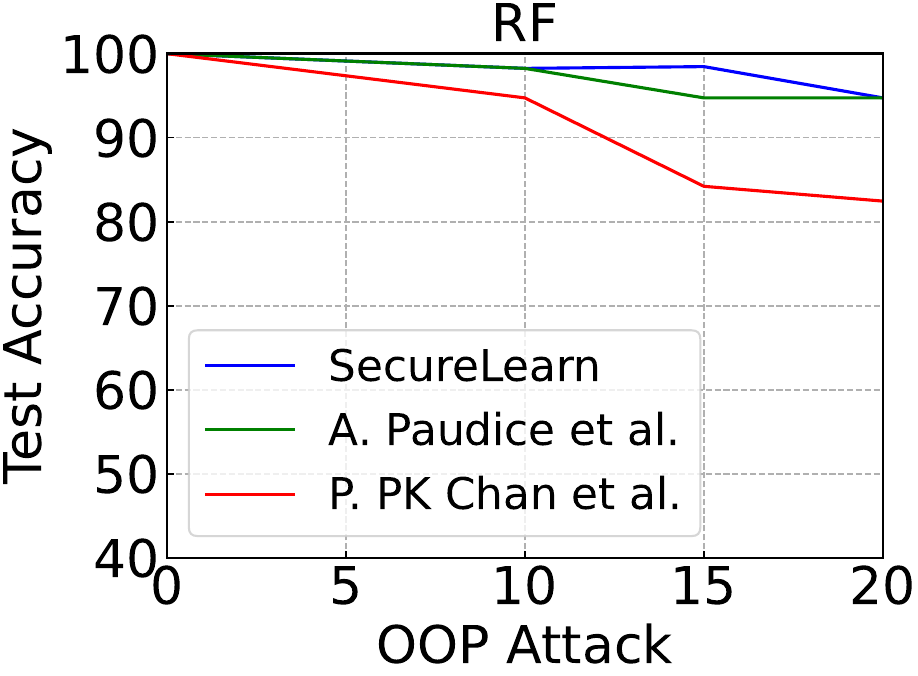}}
\subfigure
    {\includegraphics[width=3cm,height=3cm]{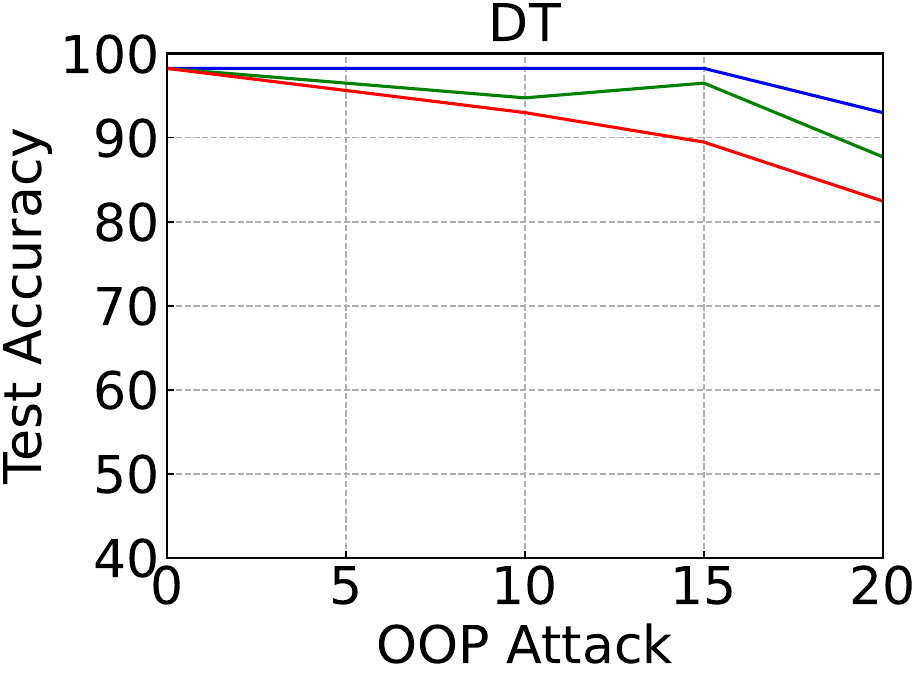}}
\subfigure
    {\includegraphics[width=3cm,height=3cm]{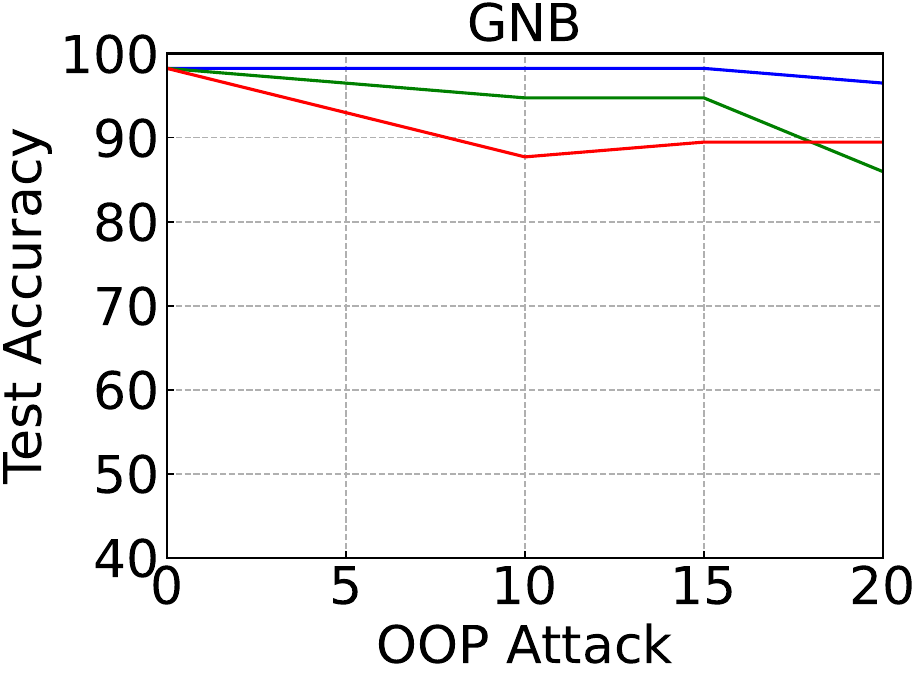}}
\subfigure
    {\includegraphics[width=3cm,height=3cm]{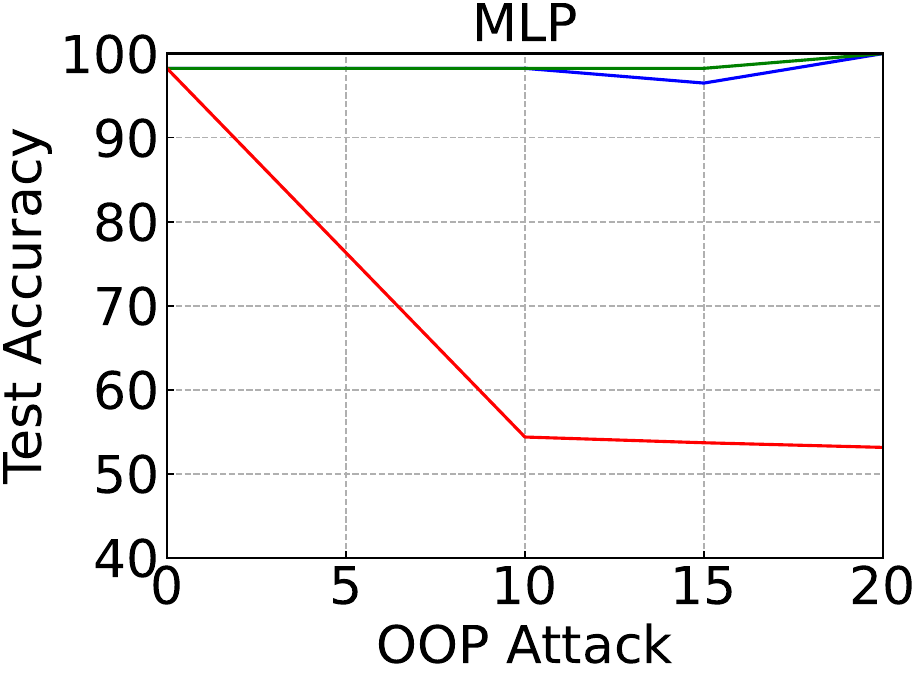}}
    \\
\subfigure
    {\includegraphics[width=3cm,height=3cm]{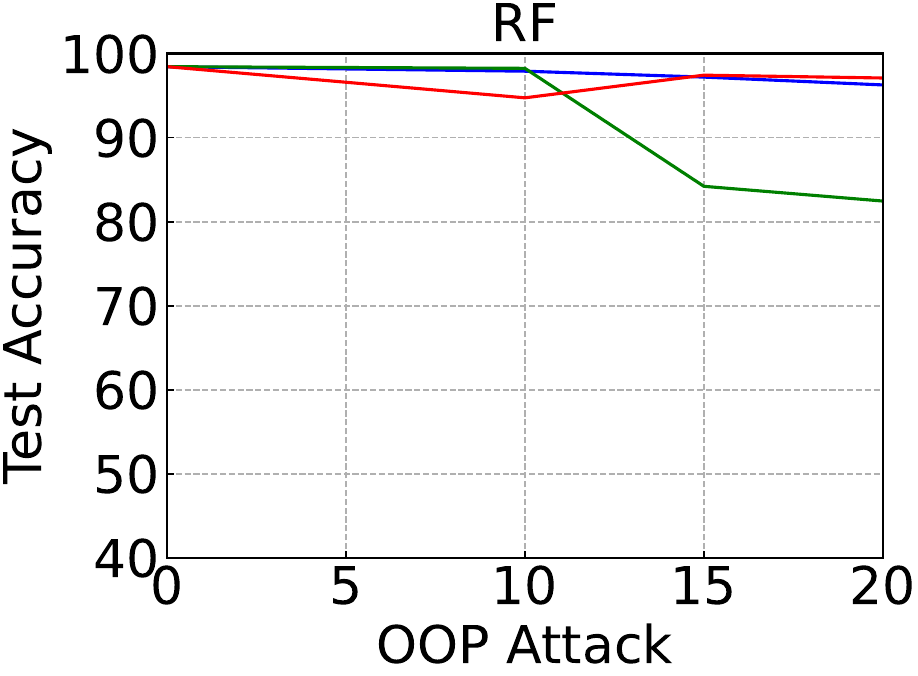}}
\subfigure
    {\includegraphics[width=3cm,height=3cm]{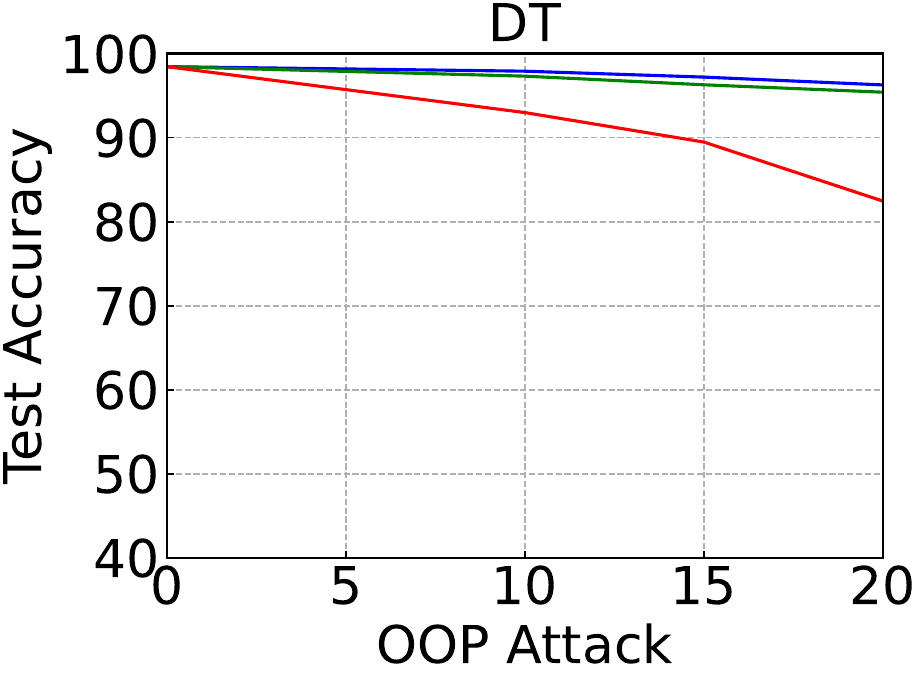}}
\subfigure
    {\includegraphics[width=3cm,height=3cm]{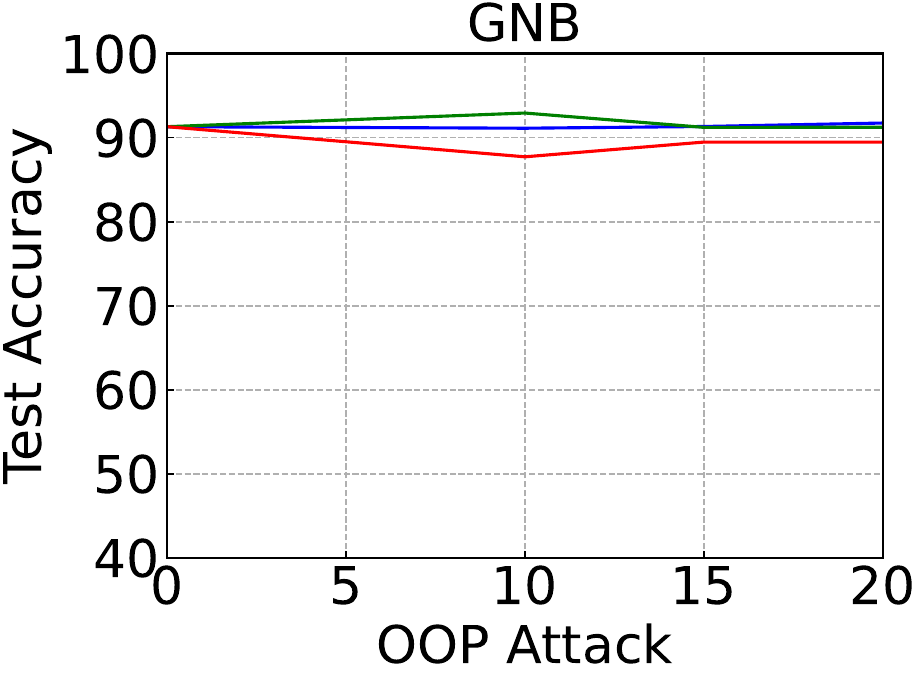}}
\subfigure
    {\includegraphics[width=3cm,height=3cm]{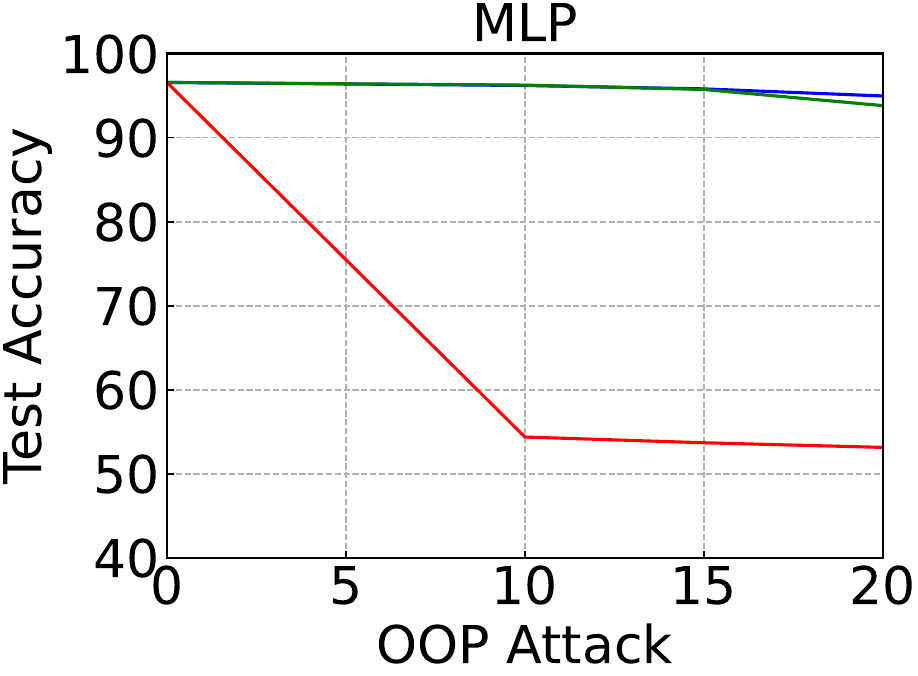}}
    \\
\subfigure
    {\includegraphics[width=3cm,height=3cm]{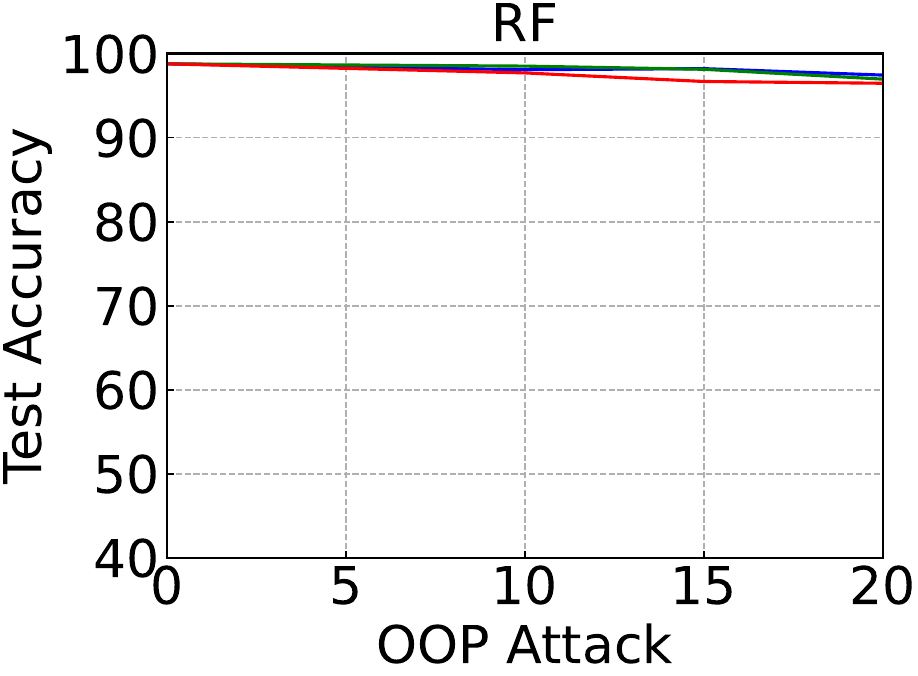}}
\subfigure
    {\includegraphics[width=3cm,height=3cm]{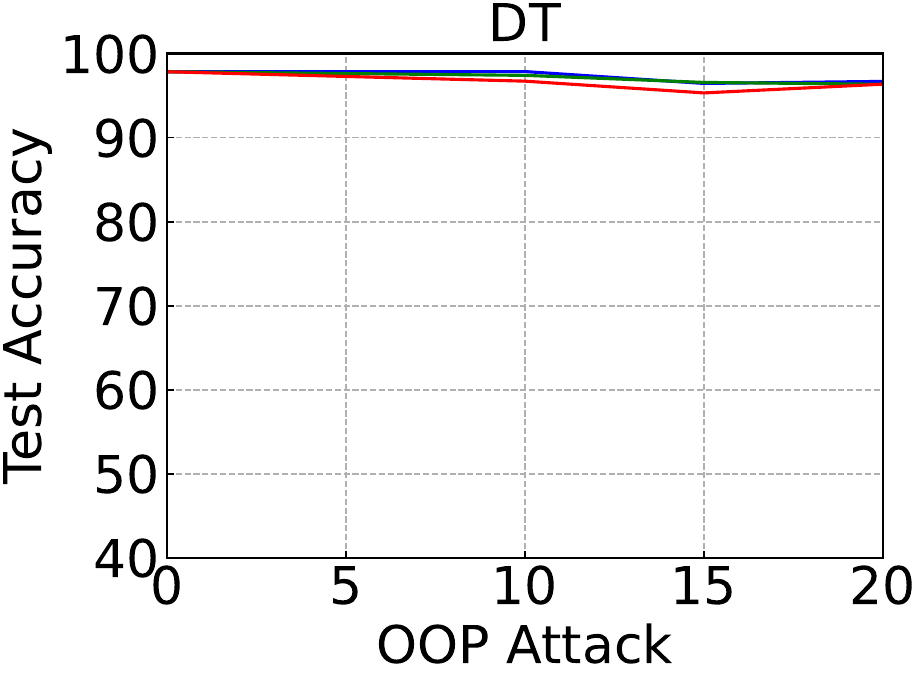}}
\subfigure
    {\includegraphics[width=3cm,height=3cm]{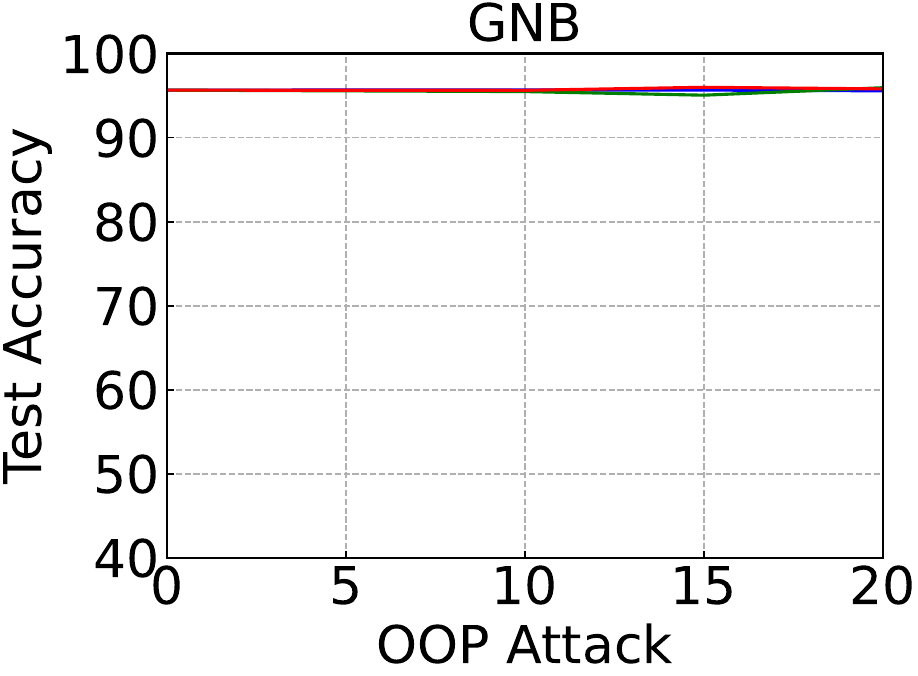}}
\subfigure
    {\includegraphics[width=3cm,height=3cm]{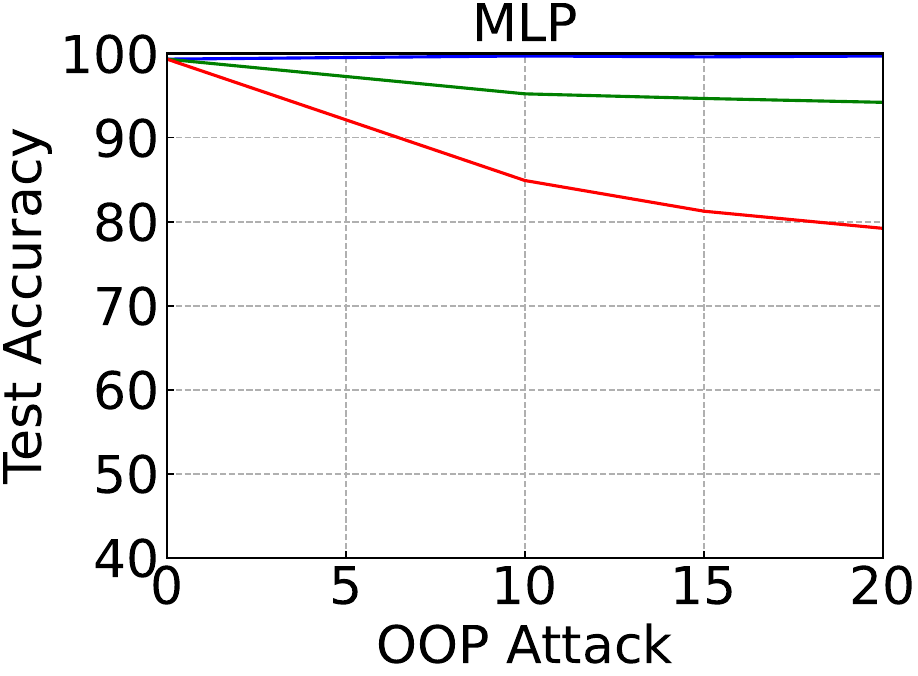}}
\caption{Impact of OOP attack on accuracy at various poisoning levels. The first row illustrates all models trained with the IRIS dataset, the models in the second row are trained with the MNIST dataset, and in the third row, the models are trained with the USPS dataset}
\label{fig:data poisoning with oop attack}
\end{figure*}

\begin{figure*}[!t]
\centering
\subfigure
    {\includegraphics[width=3cm,height=3cm]{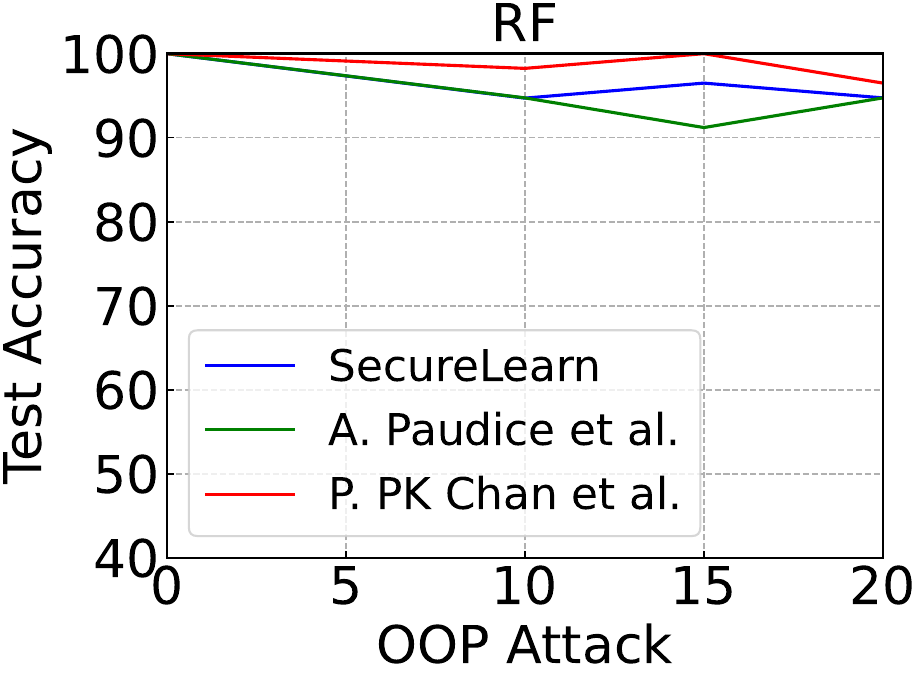}}
\subfigure
    {\includegraphics[width=3cm,height=3cm]{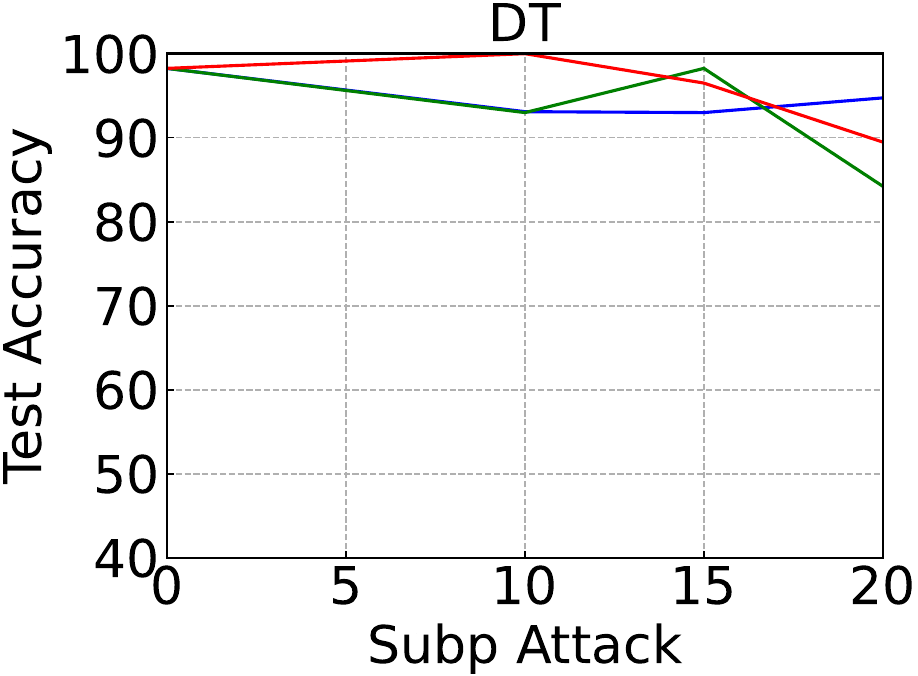}}
\subfigure
    {\includegraphics[width=3cm,height=3cm]{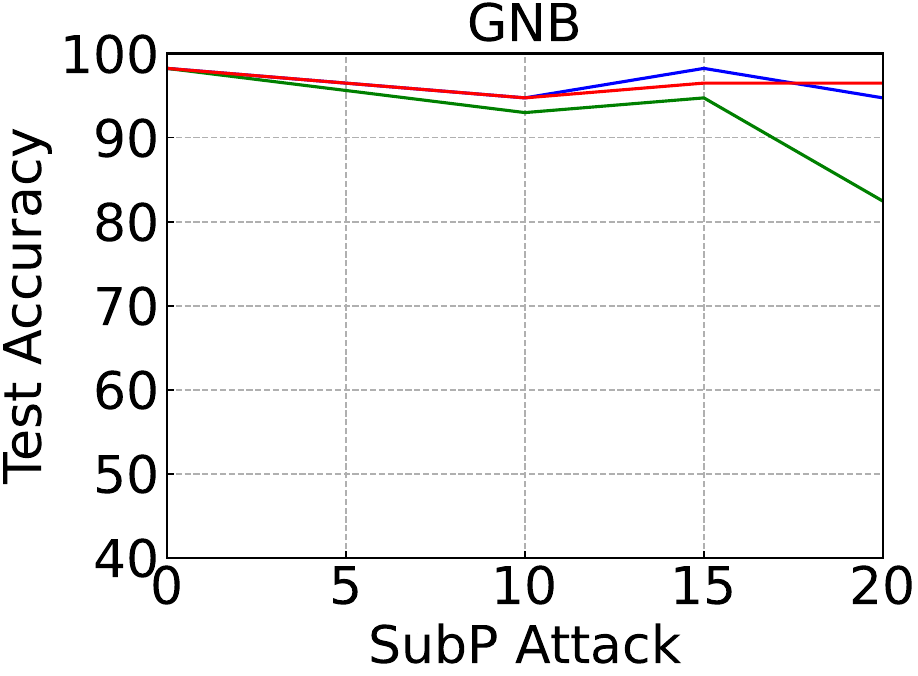}}
\subfigure
    {\includegraphics[width=3cm,height=3cm]{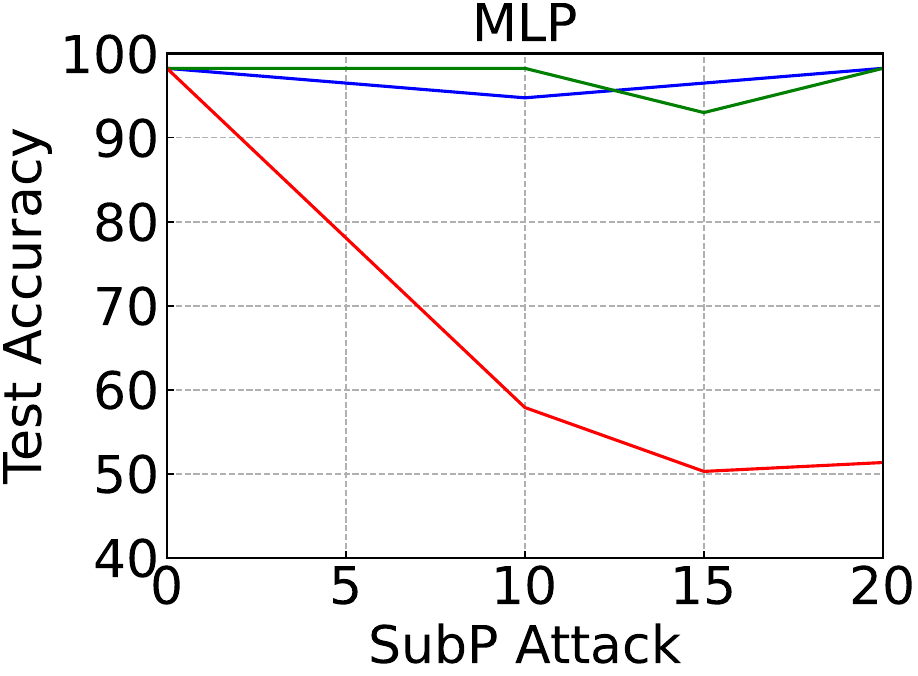}}
     \\
\subfigure
    {\includegraphics[width=3cm,height=3cm]{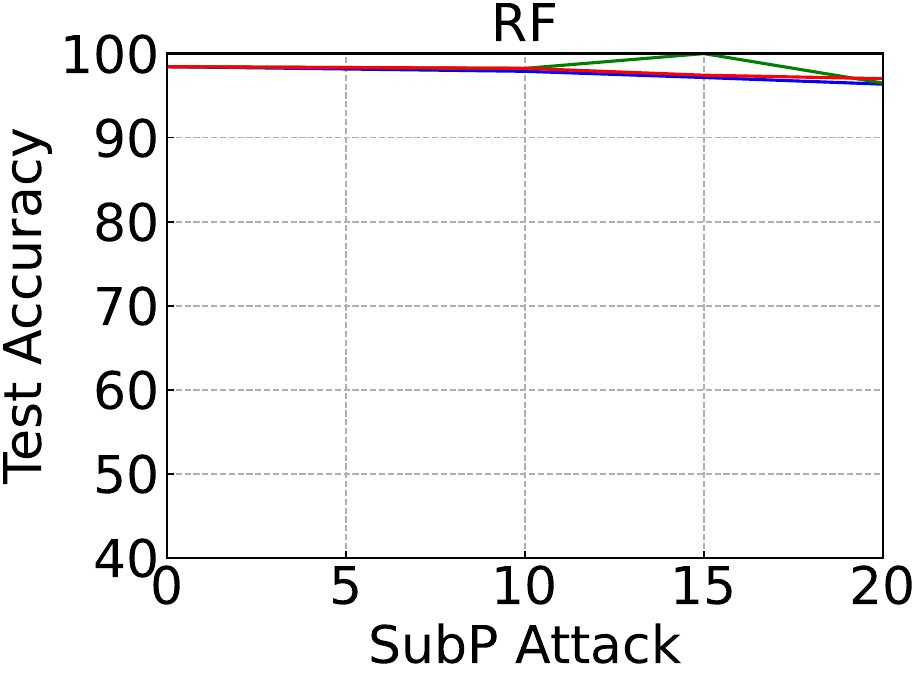}}
\subfigure
    {\includegraphics[width=3cm,height=3cm]{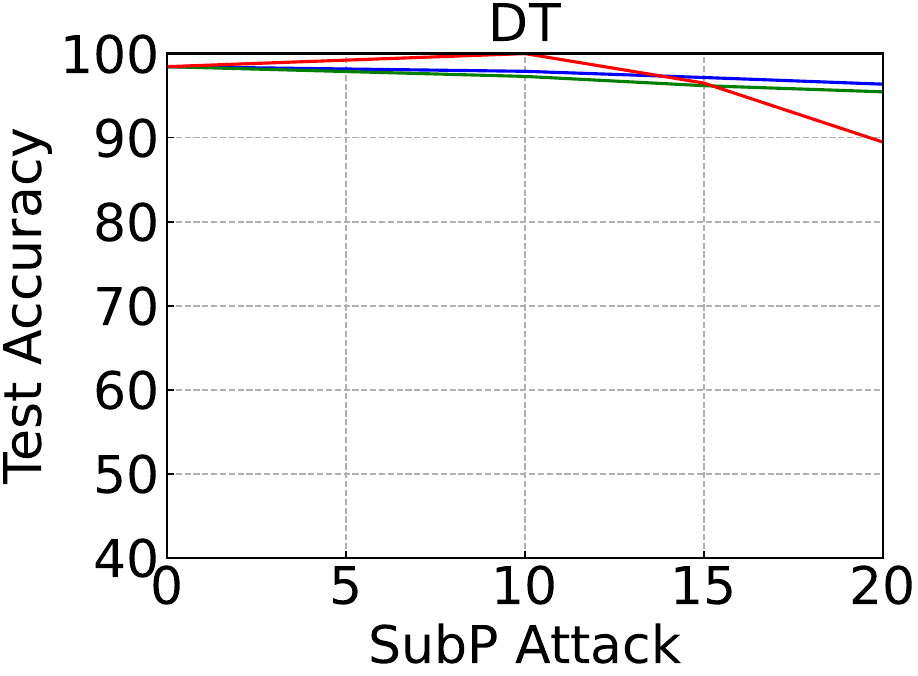}}
\subfigure
    {\includegraphics[width=3cm,height=3cm]{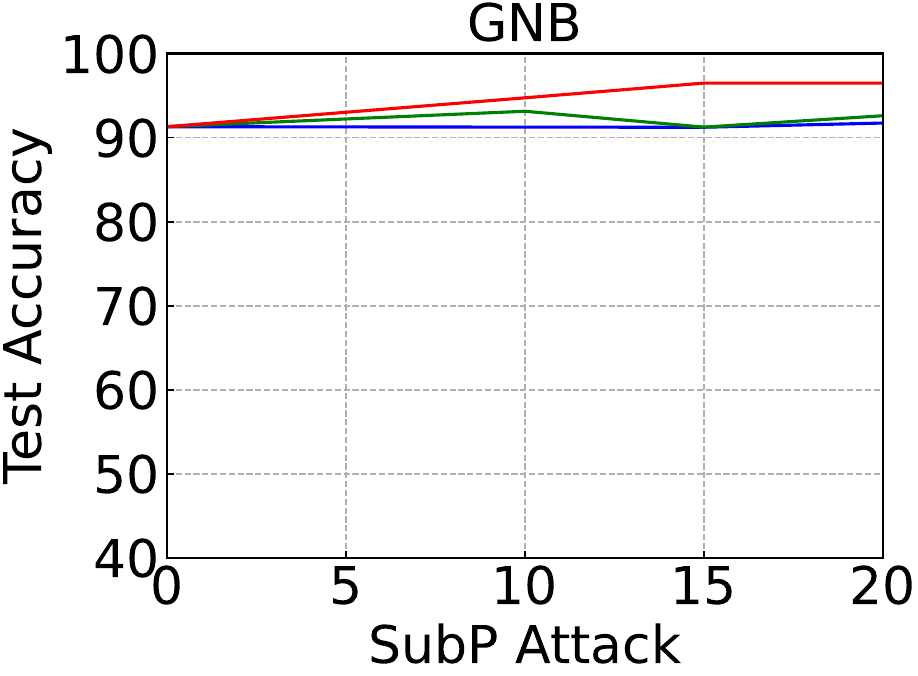}}
\subfigure
    {\includegraphics[width=3cm,height=3cm]{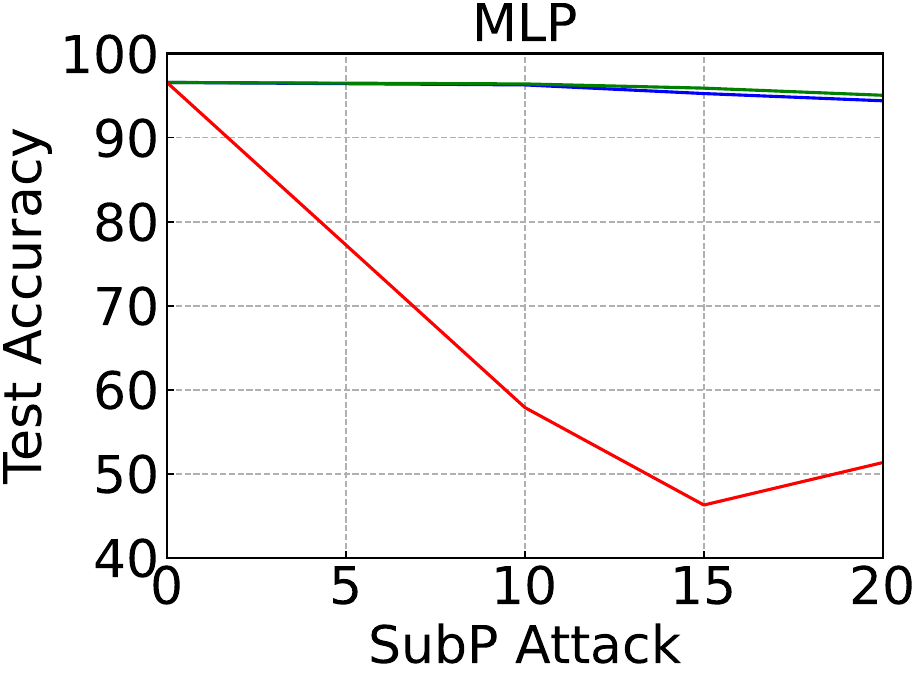}}
    \\
\subfigure
    {\includegraphics[width=3cm,height=3cm]{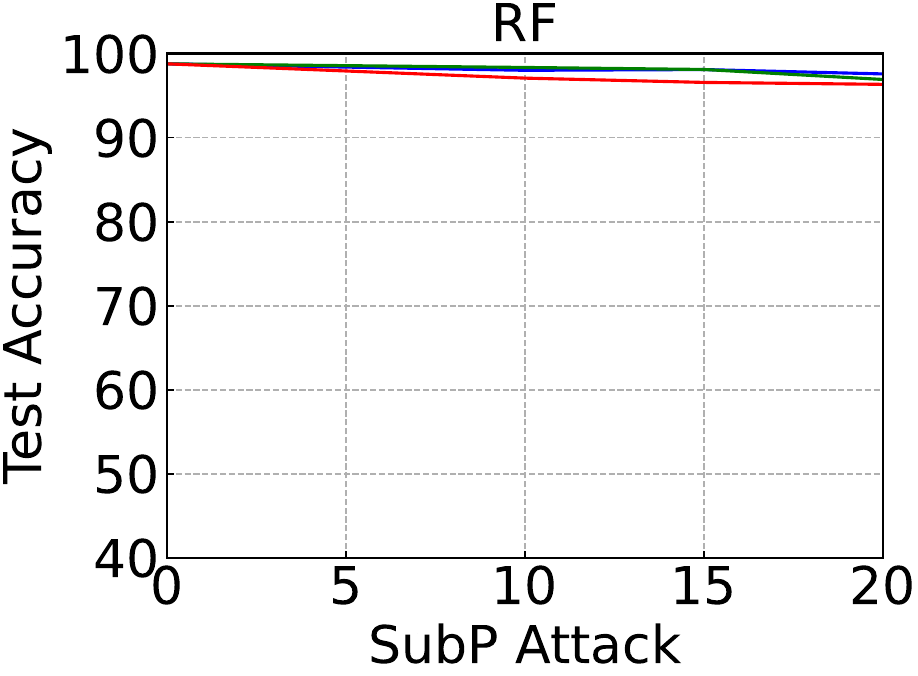}}
\subfigure
    {\includegraphics[width=3cm,height=3cm]{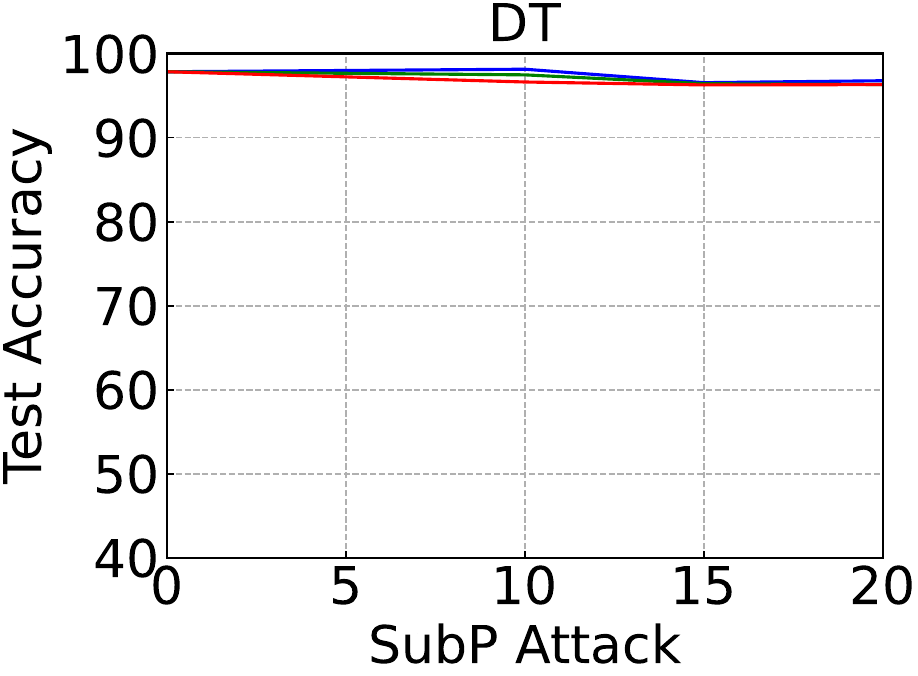}}
\subfigure
    {\includegraphics[width=3cm,height=3cm]{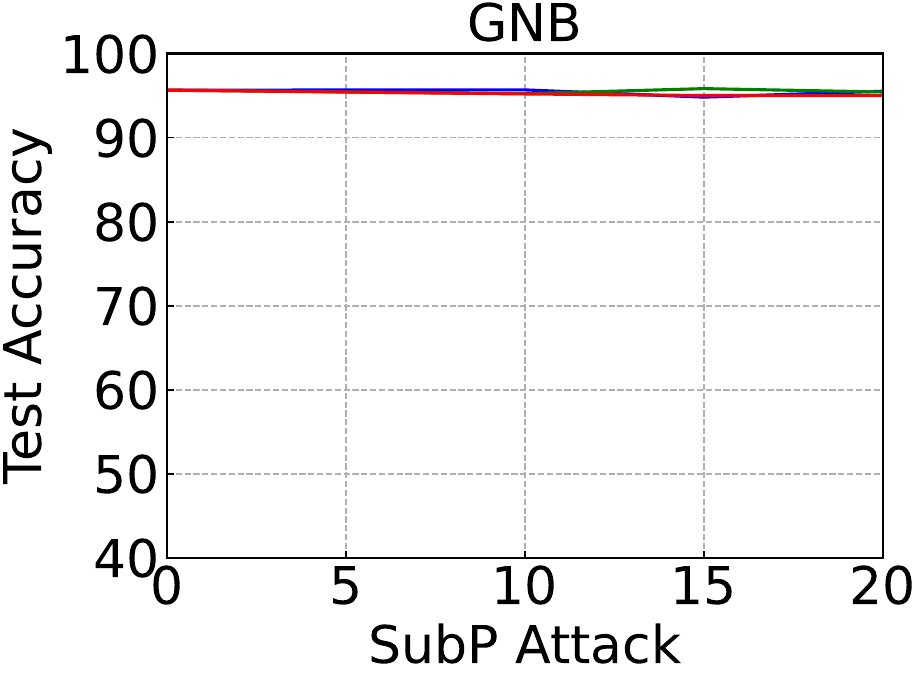}}
\subfigure
    {\includegraphics[width=3cm,height=3cm]{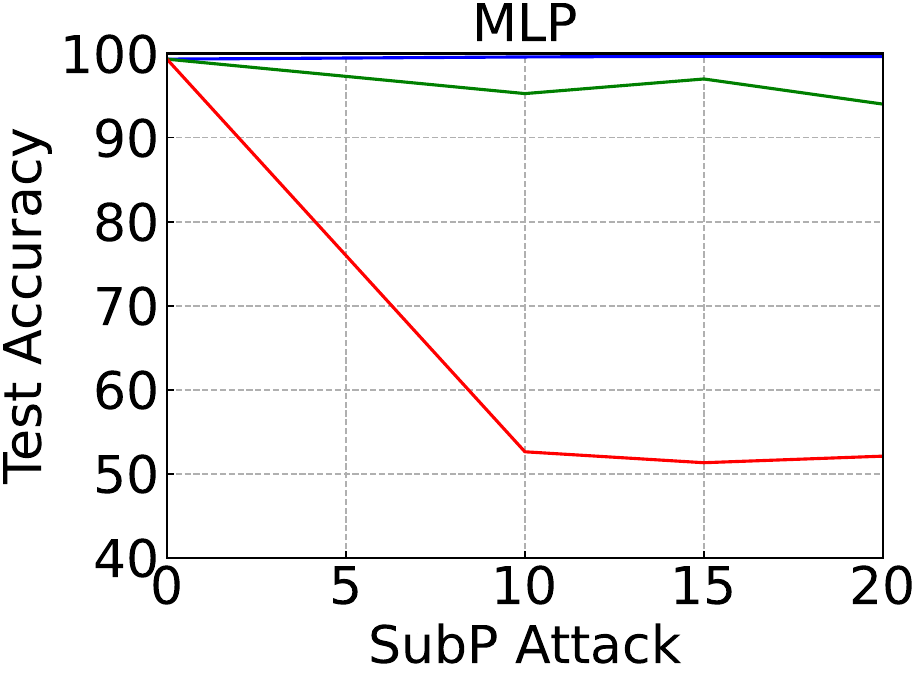}}
\caption{Impact of SubP attack on accuracy at various poisoning levels. The first row illustrates models trained with the IRIS dataset, the models in the second row are trained with the MNIST dataset, and the models in the third row are trained with the USPS dataset}
\label{fig:data poisoning with subp attack}
\end{figure*}

\begin{figure*}[!t]
\centering
\subfigure
    {\includegraphics[width=3cm,height=3cm]{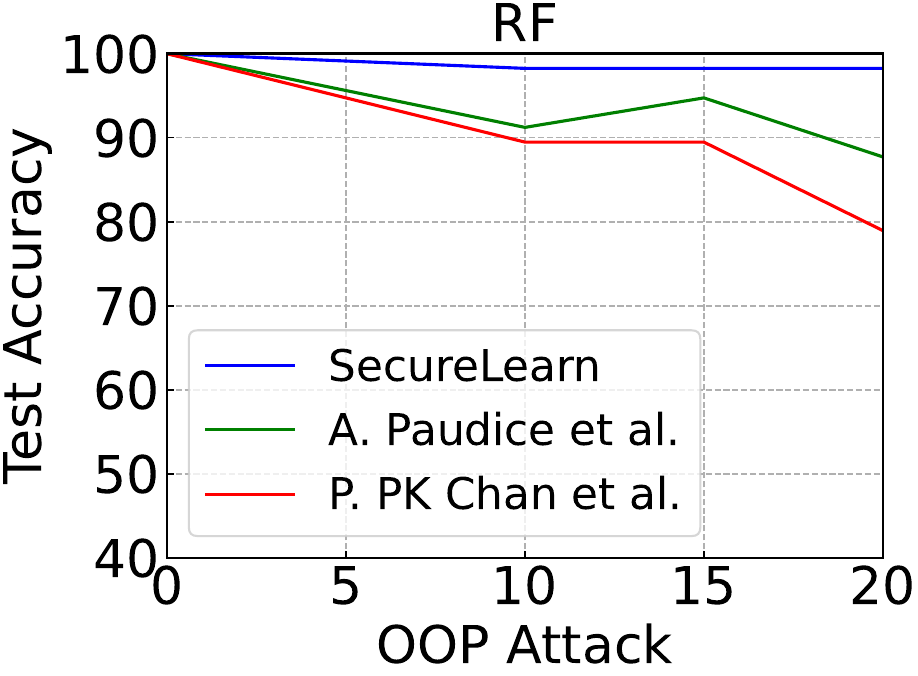}}
\subfigure
    {\includegraphics[width=3cm,height=3cm]{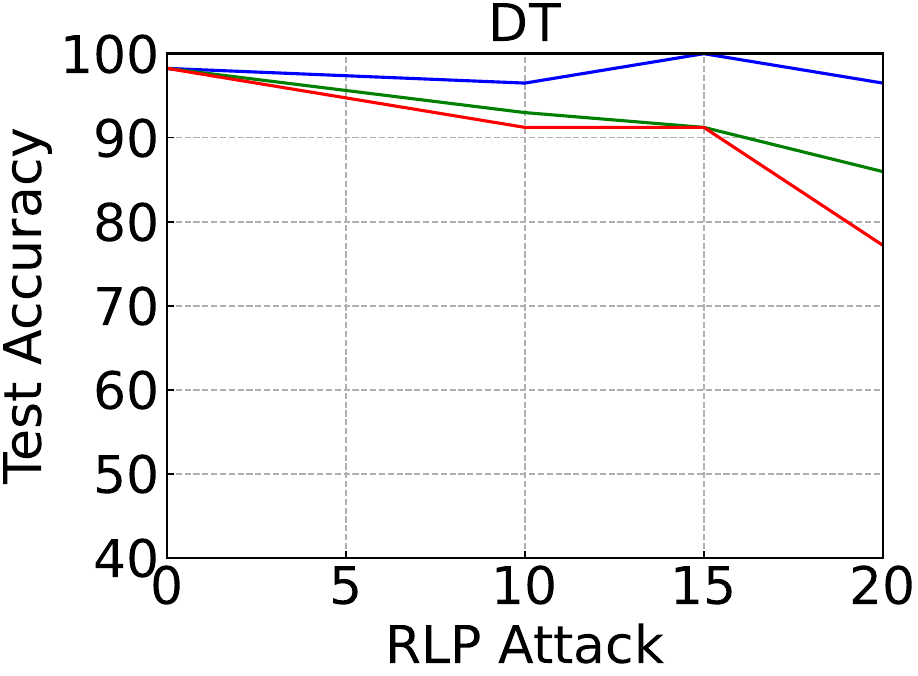}}
\subfigure
    {\includegraphics[width=3cm,height=3cm]{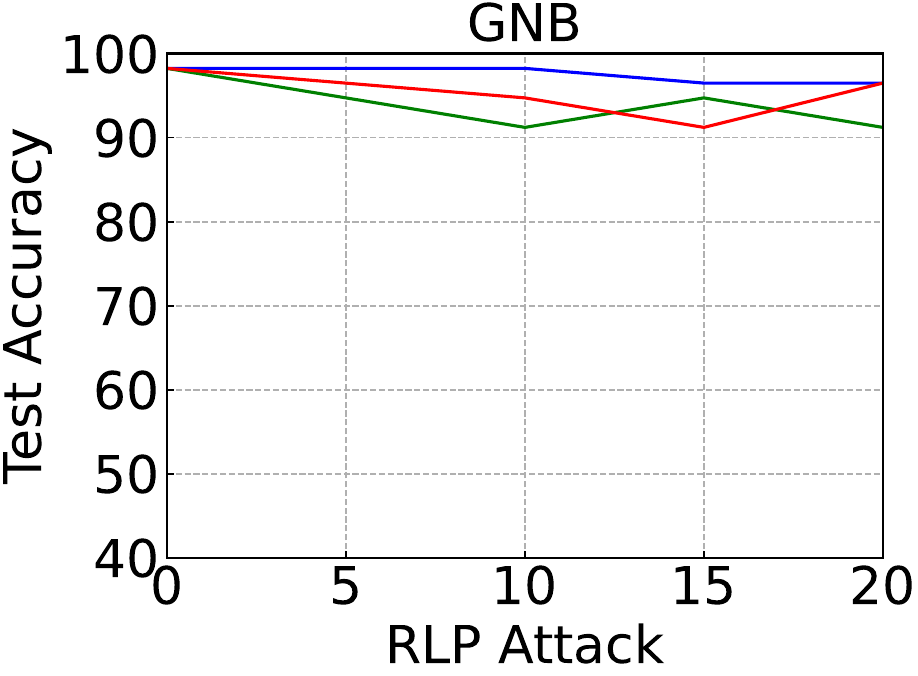}}
\subfigure
    {\includegraphics[width=3cm,height=3cm]{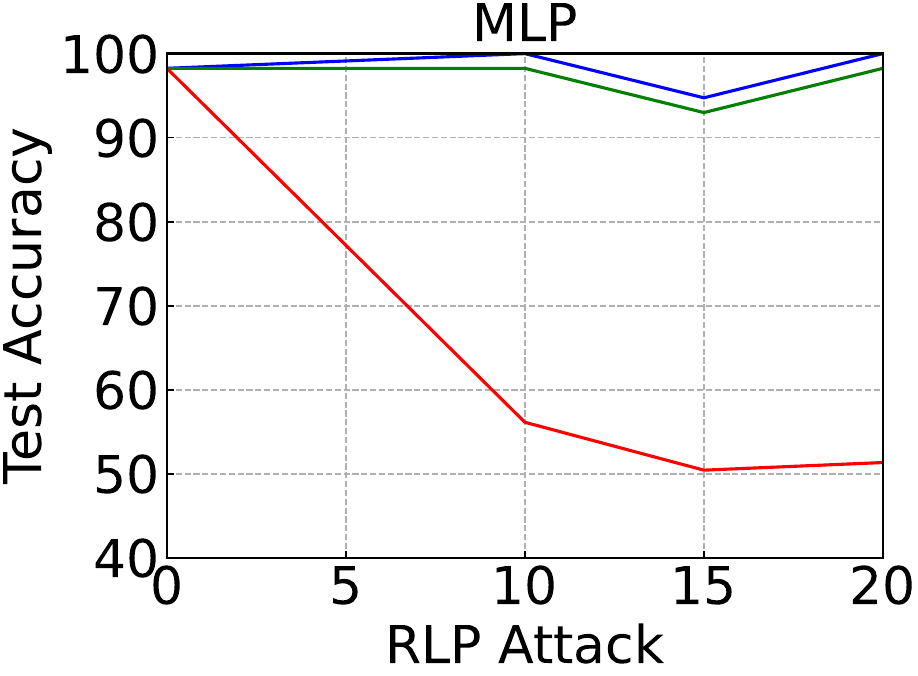}}
     \\
\subfigure
    {\includegraphics[width=3cm,height=3cm]{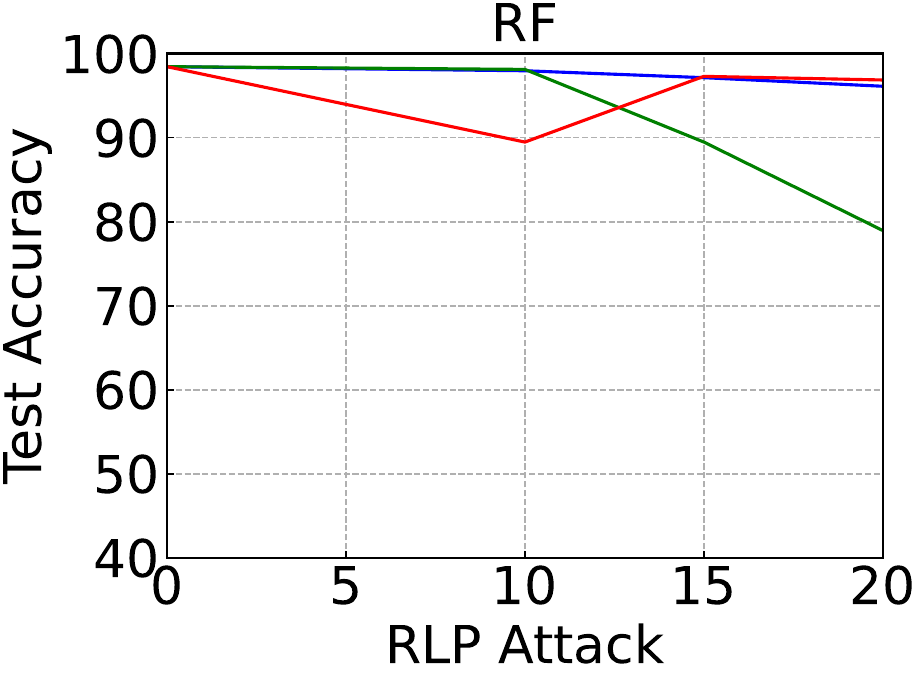}}
\subfigure
    {\includegraphics[width=3cm,height=3cm]{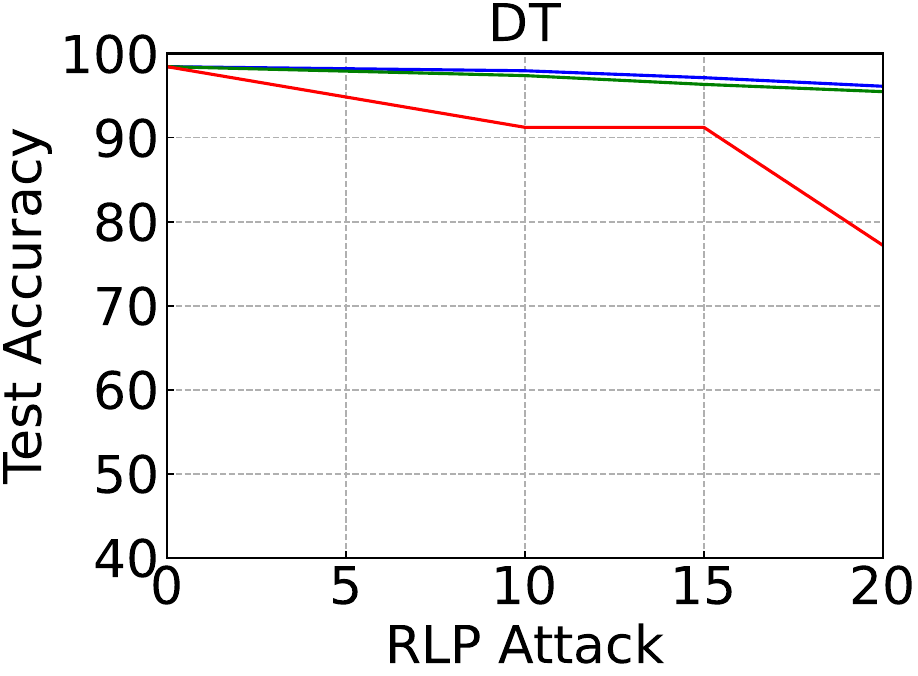}}
\subfigure
    {\includegraphics[width=3cm,height=3cm]{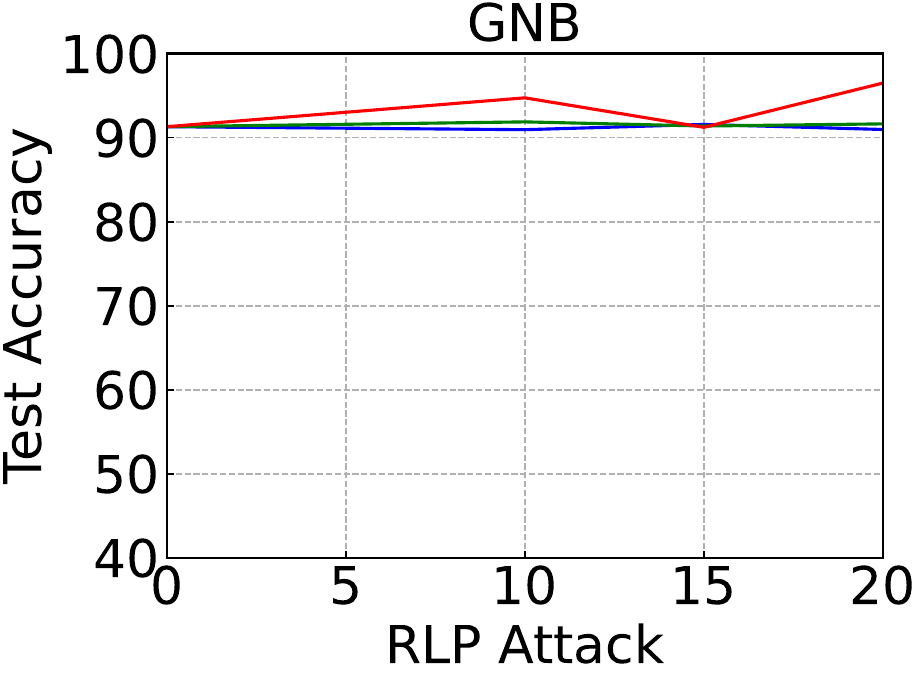}}
\subfigure
    {\includegraphics[width=3cm,height=3cm]{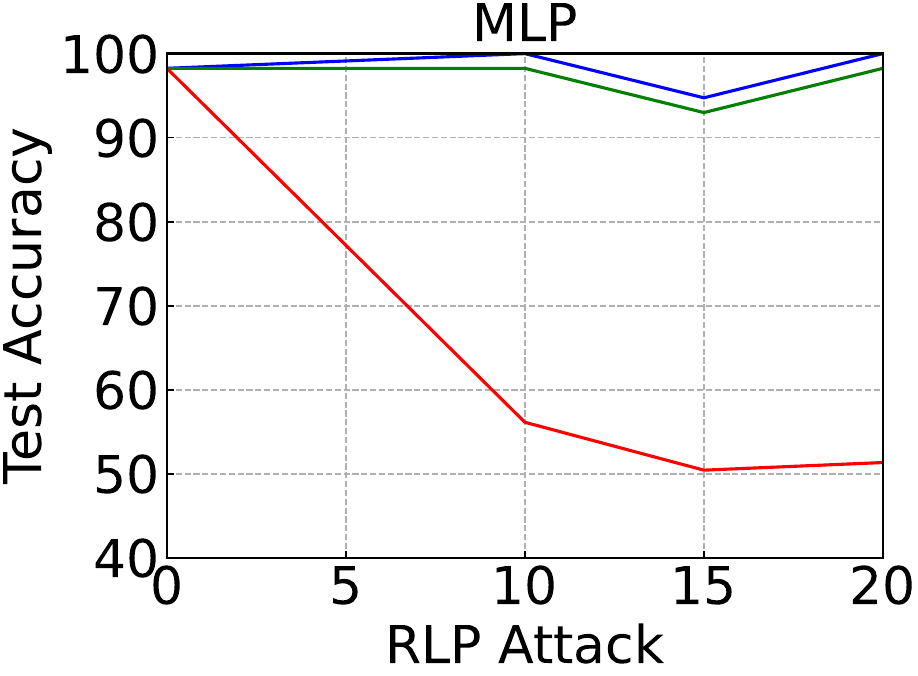}}
    \\
\subfigure
    {\includegraphics[width=3cm,height=3cm]{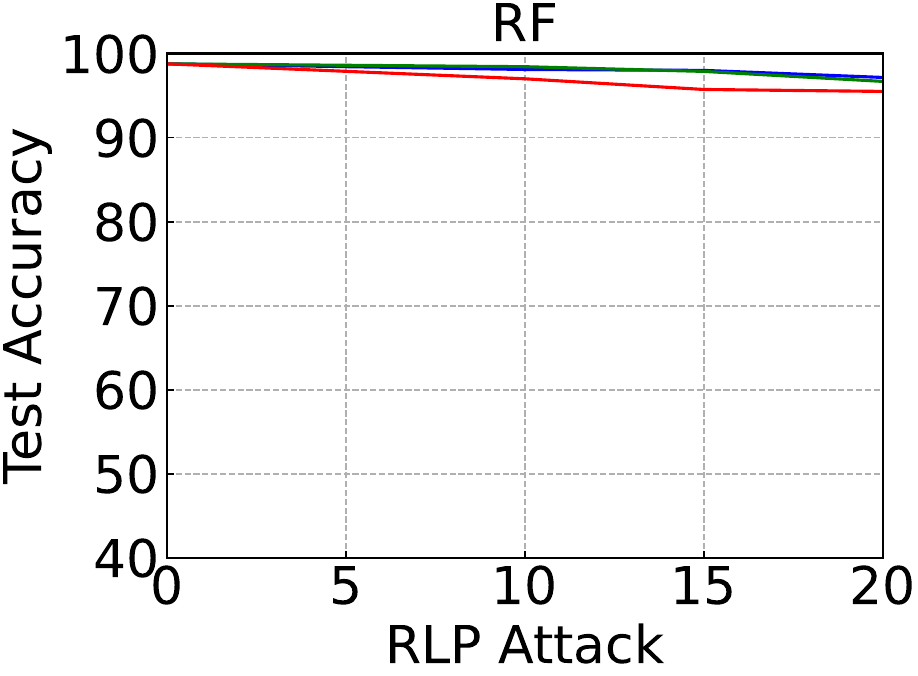}}
\subfigure
    {\includegraphics[width=3cm,height=3cm]{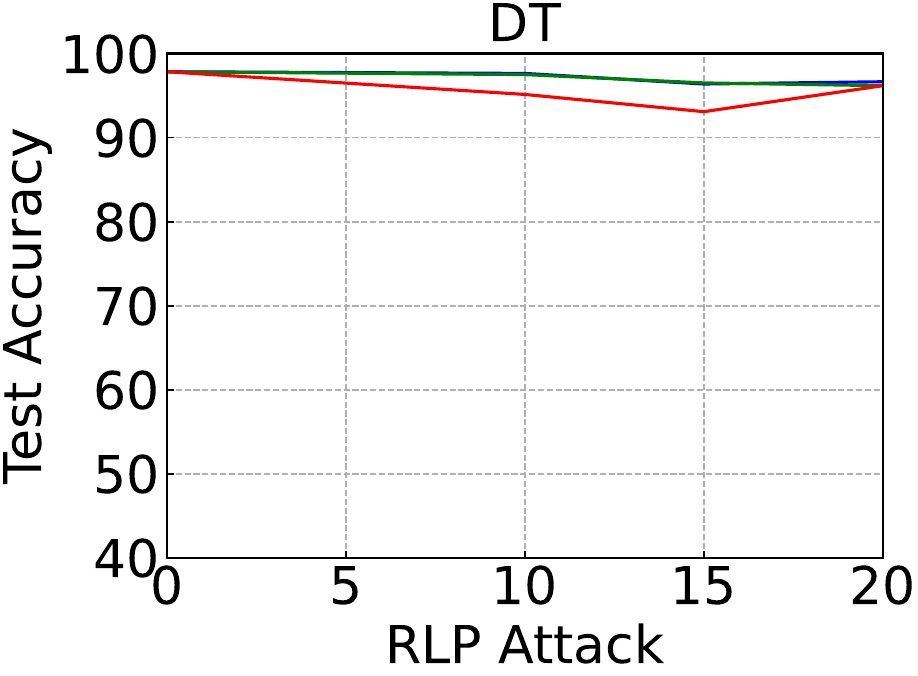}}
\subfigure
    {\includegraphics[width=3cm,height=3cm]{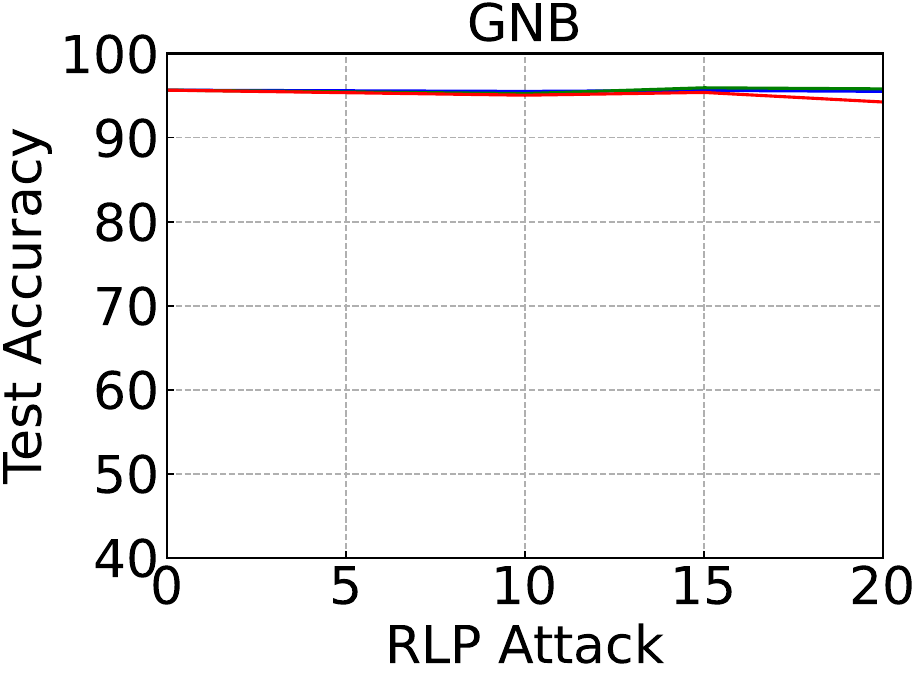}}
\subfigure
    {\includegraphics[width=3cm,height=3cm]{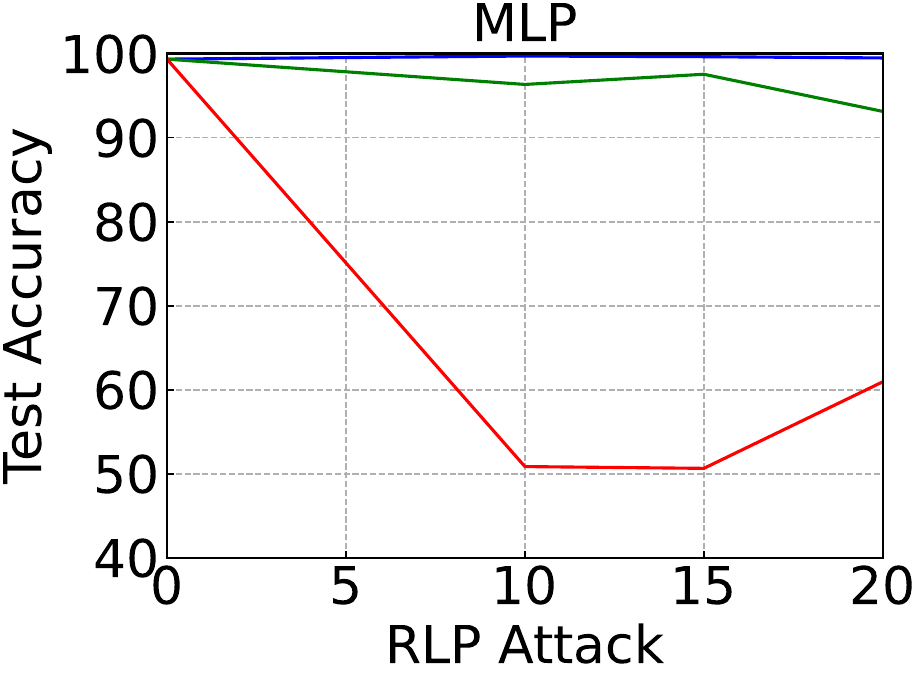}}
\caption{Impact of RLP attack on accuracy at various poisoning levels. The first row illustrates models trained with the IRIS dataset, the models in the second row are trained with the MNIST dataset, and the models in the third row are trained with the USPS dataset}
\label{fig:data poisoning with rlp attack}
\end{figure*}

We have analyzed model performance from Eq. \ref{Eq:accuracy} to Eq. \ref{Eq:f1_score} while setting the poisoning level at $10\%<\Delta L <20\%$. Baseline accuracy of models is given in Fig. \ref{fig:data poisoning with oop attack} to Fig. \ref{fig:data poisoning with rlp attack} where $\Delta L=15\%$. Our findings indicate that the data sanitization with SecureLearn outperforms other solutions and provides stable accuracy of at least 90\% across implemented data poisoning attacks. The recall and F1-score are provided in Table \ref{tab:recall_f1_score}. \\
SecureLearn outperformed the mitigations proposed in \cite{paudice2019label} and \cite{chan2018data} in sanitizing poisoned datasets. Compared to SecureLearn, the data sanitization method proposed in \cite{paudice2019label} achieved similar accuracy for DT with an average of 96\%. SecureLearn provided an average recall of 84.22\% with a 3\% higher F1-score. Similarly, the average accuracy for GNB provided by \cite{paudice2019label} is 94\%, equivalent to SecureLearn; however, its recall and F1-score are 3.69\% and 3.63\% lower, respectively. Furthermore, the sanitized accuracy provided by \cite{paudice2019label} dropped to 79\% for the RLP attack and to 82\% for the OOP attack when the model is trained with the MNIST dataset. \\
The data sanitization proposed by \cite{chan2018data} is highly unstable, particularly for MLP models. The accuracy of each model consistently decreases with increasing poisoning levels. For example, the accuracy of MLP substantially decreases after 10\% poisoning, reached approximately 52\% when trained on the IRIS and MNIST datasets, and 80\% when trained on the USPS dataset. This instability arises because the method removes anomalous data points, which potentially decreases model accuracy. However, removing such data points also reduces the dataset size, which leads to underfitting, particularly in MLP. 

\begin{table*}[!ht]
\tiny
    \centering
    \caption{Impact of data poisoning on recall and F1-score of the model}
    \label{tab:recall_f1_score}
    \resizebox{\textwidth}{!}{\begin{tabular}{c c c c c c c c c c c c c}
\toprule
\multirow{3}{*}{Metric} & \multirow{3}{*}{Algorithm} & \multirow{3}{*}{Dataset} & \multirow{3}{*}{Defense} & \multicolumn{9}{c}{Attack}\\
\cline{5-13}
& & & & \multicolumn{3}{c }{OOP} & \multicolumn{3}{ c }{Subp} & \multicolumn{3}{ c}{RLP}\\
\cline{5-7} \cline{8-10} \cline{11-13}
& & & & $\Delta L=10\% $ & $\Delta L=15\% $ & $\Delta L=20\% $ & $\Delta L=10\% $ & $\Delta L=15\% $ & $\Delta L=20\% $ & $\Delta L=10\% $ & $\Delta L=15\% $ & $\Delta L=20\% $ \\
\midrule
\multirow{36}{*}{Recall} & \multirow{9}{*}{RF} & \multirow{3}{*}{IRIS} & A. Paudice et al. & \textbf{97.33} & 92.85 & \textbf{91.66} & 91.88 & 88.09 & 91.66 & 87.17 & 92.85 & 80.55 \\
& & & M. Barreno et al. & 92.09 & 78.57 & 75.04 & \textbf{97.43} & 99.99 & \textbf{96.07} & 84.61 & 84.12 & 69.75 \\
& & & SecureLearn & 93.73 & \textbf{99.99} & 86.53 & 94.87 & \textbf{99.99} & 91.88 & \textbf{94.87} & \textbf{96.96} & \textbf{92.09} \\
& & \multirow{3}{*}{MNIST} & A. Paudice et al. & 88.22 & 85.20 & 82.19 & 88.13 & 85.44 & 82.38 & 88.57 & 85.20 & 81.96 \\
& & & M. Barreno et al. & \textbf{92.09} & 78.57 & 75.04 & \textbf{97.43} & \textbf{99.99} & \textbf{96.07} & 84.61 & 84.12 & 69.75 \\
& & & SecureLearn & 91.31 & \textbf{86.63} & \textbf{84.19} & 91.34 & 86.61 & 84.38 & \textbf{90.76} & \textbf{86.30} & \textbf{83.40} \\
& & \multirow{3}{*}{USPS} & A. Paudice et al. & 91.48 & 89.08 & 81.51 & 90.57 & 88.95 & 81.38 & 91.06 & 87.65 & 80.07 \\
& & & M. Barreno et al. & 86.84 & 81.14 & 80.50 & 83.26 & 80.35 & 80.40 & 82.85 & 75.96 & 75.86 \\
& & & SecureLearn & \textbf{95.18} & \textbf{91.02} & \textbf{90.16} & \textbf{95.36} & \textbf{90.51} & \textbf{90.56} & \textbf{95.33} & \textbf{90.22} & \textbf{89.16} \\\cline{2-13}
& \multirow{9}{*}{DT} & \multirow{3}{*}{IRIS} & A. Paudice et al. & 93.74 & 94.88 & 82.05 & 91.66 & \textbf{97.91} & 74.64 & 85.79 & 92.85 & 79.48 \\
& & & M. Barreno et al. & 86.66 & 81.81 & 77.77 & \textbf{99.99} & 93.93 & \textbf{85.18} & 84.70 & 84.84 & 69.62 \\
& & & SecureLearn & \textbf{97.77} & \textbf{97.91} & \textbf{88.88} & 95.55 & 94.21 & 84.12 & \textbf{95.55} & \textbf{94.21} & \textbf{83.33} \\
& & \multirow{3}{*}{MNIST} & A. Paudice et al. & \textbf{86.93} & 81.84 & 78.09 & 86.90 & 81.38 & 78.21 & \textbf{86.71} & 81.93 & \textbf{78.28} \\
& & & M. Barreno et al. & 86.66 & 81.81 & 77.77 & \textbf{99.99} & \textbf{93.93} & \textbf{85.18} & 84.56 & \textbf{84.84} & 69.62 \\
& & & SecureLearn & 85.45 & \textbf{85.13} & \textbf{78.20} & 85.45 & 84.38 & 78.40 & 85.45 & 84.56 & 77.44 \\
& & \multirow{3}{*}{USPS} & A. Paudice et al. & 85.67 & 80.39 & 80.63 & 86.14 & 79.60 & 80.41 & 86.27 & 80.12 & 79.61 \\
& & & M. Barreno et al. & 81.65 & 74.01 & 80.29 & 81.34 & 79.41 & 80.07 & 73.85 & 62.92 & 79.25 \\
& & & SecureLearn & \textbf{87.42} & \textbf{81.51} & \textbf{81.00} & \textbf{87.37} & \textbf{81.58} & \textbf{81.50} & \textbf{87.40} & \textbf{81.55} & \textbf{79.82} \\\cline{2-13}
& \multirow{9}{*}{GNB} & \multirow{3}{*}{IRIS} & A. Paudice et al. & 91.11 & \textbf{94.11} & 77.77 & 88.88 & 94.11 & 71.96 & 86.11 & 94.11 & 86.11 \\
& & & M. Barreno et al. & 85.18 & 84.40 & 85.30 & 92.59 & \textbf{94.65} & 94.74 & 90.74 & 86.96 & 94.74 \\
& & & SecureLearn & \textbf{95.39} & 92.59 & \textbf{98.03} & \textbf{95.39} & 94.44 & \textbf{98.03} & \textbf{95.39} & \textbf{94.44} & \textbf{98.03} \\
& & \multirow{3}{*}{MNIST} & A. Paudice et al. & 57.12 & 60.34 & 58.98 & 58.39 & 57.78 & 52.72 & 56.49 & 59.48 & 50.50 \\
& & & M. Barreno et al. & \textbf{85.18} & \textbf{84.40} & \textbf{85.30} & \textbf{92.59} & \textbf{94.65} & \textbf{94.74} & \textbf{90.74} & \textbf{86.96} & \textbf{94.74} \\
& & & SecureLearn & 57.71 & 57.12 & 57.15 & 57.93 & 58.38 & 57.65 & 58.48 & 57.33 & 57.16 \\
& & \multirow{3}{*}{USPS} & A. Paudice et al. & 75.39 & 73.11 & 77.28 & 74.01 & \textbf{77.54} & 75.01 & 74.01 & 77.67 & 76.64 \\
& & & M. Barreno et al. & 76.70 & 75.94 & 76.19 & 76.85 & 71.12 & 75.82 & 75.73 & 75.94 & 75.83 \\
& & & SecureLearn & \textbf{76.97} & \textbf{78.16} & \textbf{77.50} & \textbf{77.34} & 76.80 & \textbf{77.23} & \textbf{76.57} & \textbf{77.93} & \textbf{77.26} \\\cline{2-13}
& \multirow{9}{*}{MLP} & \multirow{3}{*}{IRIS} & A. Paudice et al. & 96.29 & 97.77 & \textbf{99.99} & 96.27 & 91.11 & \textbf{97.22} & 96.3 & 90.47 & 97.22 \\
& & & M. Barreno et al. & 31.11 & 28.61 & 36.01 & 36.30 & 18.72 & 28.51 & 33.92 & 16.34 & 28.51 \\
& & & SecureLearn & \textbf{99.90} & \textbf{98.01} & 99.90 & \textbf{99.99} & \textbf{97.98} & 96.96 & \textbf{99.99} & \textbf{99.90} & \textbf{99.99} \\
& & \multirow{3}{*}{MNIST} & A. Paudice et al. & 96.29 & \textbf{97.77} & \textbf{99.99} & 96.15 & 91.11 & 97.22 & 96.29 & 90.47 & 97.22 \\
& & & M. Barreno et al. & 31.11 & 28.61 & 36.01 & 36.30 & 18.72 & 28.51 & 33.92 & 16.34 & 28.51 \\
& & & SecureLearn & \textbf{97.93} & 97.45 & 97.05 & \textbf{98.08} & \textbf{97.82} & \textbf{97.37} & \textbf{97.32} & \textbf{97.60} & \textbf{97.25} \\
& & \multirow{3}{*}{USPS} & A. Paudice et al. & 96.29 & 82.92 & 83.52 & 96.30 & 81.05 & 79.69 & 96.29 & 81.04 & 82.33 \\
& & & M. Barreno et al. & 85.56 & 78.9 & 83.52 & 86.10 & 51.47 & 79.69 & 86.04 & 82.33 & 79.20 \\
& & & SecureLearn & \textbf{98.42} & \textbf{97.76} & \textbf{98.40} & \textbf{97.69} & \textbf{98.19} & \textbf{98.05} & \textbf{98.36} & \textbf{97.87} & \textbf{97.06} \\\hline\hline
\multirow{36}{*}{F1-Score} & \multirow{9}{*}{RF} & \multirow{3}{*}{IRIS} & A. Paudice et al. & \textbf{97.33} & 91.81 & \textbf{91.65} & 91.93 & 86.49 & 91.72 & 86.06 & 91.65 & 80.37 \\
& & & M. Barreno et al. & 91.98 & 75.94 & 72.38 & \textbf{97.33} & 99.99 & \textbf{95.13} & 83.59 & 84.12 & 68.05 \\
& & & SecureLearn & 93.73 & \textbf{99.99} & 86.58 & 93.88 & \textbf{99.99} & 91.94 & \textbf{93.88} & \textbf{97.40} & \textbf{91.98} \\
& & \multirow{2}{*}{MNIST} & A. Paudice et al. & 86.05 & 82.90 & 78.60 & 85.94 & 83.13 & 78.88 & 86.27 & 82.91 & 78.40 \\
& & & M. Barreno et al. & \textbf{91.98} & 75.94 & 72.38 & \textbf{97.33} & \textbf{99.99} & \textbf{95.13} & 83.59 & \textbf{84.12} & 68.05 \\
& & & SecureLearn & 90.90 & \textbf{84.46} & \textbf{81.54} & 90.91 & 84.39 & 81.78 & \textbf{90.31} & 84.11 & \textbf{80.65} \\
& & \multirow{2}{*}{USPS} & A. Paudice et al. & 91.36 & 88.60 & 80.65 & 90.45 & 88.53 & 80.49 & 91.00 & 87.18 & 78.82 \\
& & & M. Barreno et al. & 86.47 & 79.05 & 79.47 & 83.26 & 78.42 & 79.23 & 82.85 & 74.38 & 74.84 \\
& & & SecureLearn & \textbf{95.17} & \textbf{90.85} & \textbf{88.94} & \textbf{95.36} & \textbf{90.44} & \textbf{89.34} & \textbf{95.26} & \textbf{90.09} & \textbf{87.96} \\\cline{2-13}
& \multirow{9}{*}{DT} & \multirow{2}{*}{IRIS} & A. Paudice et al. & 93.52 & 94.88 & 78.80 & 91.31 & \textbf{97.47} & 73.68 & 89.98 & 85.85 & 75.42 \\
& & & M. Barreno et al. & 88.15 & 81.56 & 70.85 & \textbf{99.99} & 93.88 & 83.81 & 84.56 & 84.84 & 61.16 \\
& & & SecureLearn & \textbf{97.77} & \textbf{97.16} & \textbf{89.16} & 94.66 & 94.21 & \textbf{83.82} & \textbf{95.53} & \textbf{94.21} & \textbf{82.50} \\
& & \multirow{2}{*}{MNIST} & A. Paudice et al. & 86.38 & 79.11 & 75.56 & 86.27 & 78.61 & 75.50 & \textbf{86.08} & 79.33 & \textbf{75.75} \\
& & & M. Barreno et al. & \textbf{88.15} & 81.56 & 70.85 & \textbf{99.99} & \textbf{93.88} & \textbf{83.81} & 84.56 & \textbf{84.81} & 61.16 \\
& & & SecureLearn & 84.70 & \textbf{84.58} & \textbf{75.80} & 84.70 & 83.70 & 75.85 & 84.52 & 83.89 & 74.49 \\
& & \multirow{2}{*}{USPS} & A. Paudice et al. & 83.12 & 77.46 & 77.40 & 83.48 & 76.58 & 77.03 & 83.79 & 77.18 & 76.06 \\
& & & M. Barreno et al. & 79.78 & 70.65 & 76.71 & 79.24 & 76.10 & 76.42 & 71.73 & 60.37 & 75.67 \\
& & & SecureLearn & \textbf{85.09} & \textbf{78.66} & \textbf{81.00} & \textbf{84.97} & \textbf{78.82} & \textbf{81.50} & \textbf{84.77} & \textbf{81.55} & \textbf{78.70} \\\cline{2-13}
& \multirow{9}{*}{GNB} & \multirow{2}{*}{IRIS} & A. Paudice et al. & 90.89 & \textbf{92.77} & 76.31 & 87.77 & 92.77 & 69.88 & 84.56 & 92.77 & 86.46 \\
& & & M. Barreno et al. & 82.32 & 83.76 & 84.74 & 91.87 & \textbf{94.75} & 94.74 & 91.41 & 86.58 & 94.74 \\
& & & SecureLearn & \textbf{95.39} & 91.87 & \textbf{97.23} & \textbf{95.39} & 94.44 & \textbf{97.23} & \textbf{95.39} & \textbf{94.44} & \textbf{97.23} \\
& & \multirow{2}{*}{MNIST} & A. Paudice et al. & 53.35 & 57.86 & 56.42 & 54.81 & 55.09 & 49.86 & 52.64 & 56.94 & 49.19 \\
& & & M. Barreno et al. & \textbf{82.32} & \textbf{83.76} & \textbf{84.74} & \textbf{91.87} & \textbf{94.75} & \textbf{94.74} & \textbf{91.41} & \textbf{86.58} & \textbf{94.74} \\
& & & SecureLearn & 53.92 & 53.68 & 53.49 & 54.19 & 54.95 & 54.09 & 54.67 & 53.90 & 54.38 \\
& & \multirow{2}{*}{USPS} & A. Paudice et al. & 75.14 & 73.56 & 77.36 & 73.73 & \textbf{77.45} & 75.72 & 73.98 & \textbf{77.90} & 76.97 \\
& & & M. Barreno et al. & 76.54 & \textbf{78.37} & 76.61 & 74.27 & 71.42 & 71.45 & 72.91 & 75.19 & 66.96 \\
& & & SecureLearn & \textbf{76.79} & 77.77 & \textbf{77.47} & \textbf{77.20} & 76.62 & \textbf{77.33} & \textbf{76.55} & 77.59 & \textbf{77.14} \\\cline{2-13}
& \multirow{9}{*}{MLP} & \multirow{2}{*}{IRIS} & A. Paudice et al. & 97.18 & 97.77 & 97.70 & 97.18 & 91.11 & 90.70 & 97.18 & 90.47 & 90.52 \\
& & & M. Barreno et al. & 31.61 & 29.75 & 30.74 & 36.53 & 18.19 & 26.96 & 32.93 & 15.25 & 26.96 \\
& & & SecureLearn & \textbf{99.90} & \textbf{99.87} & \textbf{99.90} & \textbf{99.99} & \textbf{97.06} & \textbf{97.07} & \textbf{99.99} & \textbf{99.90} & \textbf{99.99} \\
& & \multirow{2}{*}{MNIST} & A. Paudice et al. & 97.18 & \textbf{97.70} & \textbf{99.99} & 97.18 & 90.70 & \textbf{97.54} & 97.18 & 90.52 & 96.96 \\
& & & M. Barreno et al. & 31.61 & 29.75 & 30.74 & 36.53 & 18.19 & 26.96 & 32.93 & 15.25 & 26.96 \\
& & & SecureLearn & \textbf{97.96} & 97.46 & 97.06 & \textbf{98.08} & \textbf{97.84} & 97.39 & \textbf{97.34} & \textbf{97.61} & \textbf{97.26} \\
& & \multirow{2}{*}{USPS} & A. Paudice et al. & 86.00 & 81.71 & 82.48 & 85.35 & 80.83 & 80.98 & 87.39 & 80.75 & 80.77 \\
& & & M. Barreno et al. & \textbf{99.99} & 78.9 & 83.52 & 14.96 & 51.47 & 79.69 & 13.88 & 82.33 & 79.20 \\
& & & SecureLearn & 98.42 & \textbf{97.77} & \textbf{98.41} & \textbf{97.74} & \textbf{98.22} & \textbf{98.08} & \textbf{98.40} & \textbf{97.95} & \textbf{97.08} \\
\bottomrule 
    \end{tabular}}
    \vspace{0.3cm}
\end{table*}


\subsubsection{\textbf{Effectiveness Of Feature-Oriented Adversarial Training (FORT)}} \label{fo_trn} We next evaluated the effectiveness of FORT in enhancing adversarial robustness of ML against data poisoning attacks. Under the same attack setting, we analyzed the change in the FDR of the model from Eq. \ref{Eq:fdr} and the results are given in Table \ref{tab:rf_fdr} to Table \ref{tab:mlp_fdr}. These results highlighted that FORT highly improved the adversarial robustness of multiclass models against all implemented data poisoning attacks. \\
These improvements are attributed to FORT's design, which leverages feature importance scores to guide adversarial training of ML. The adversarial samples for the training are developed by slightly perturbing data points close to decision boundaries and with high feature importance scores.  Generalizing over these perturbations enables the model to resist changes in its decision mechanisms with poisoned datasets. \\
The results given in the Table \ref{tab:rf_fdr} highlighted that FORT reduces the FDR of the RF model to 0.06 when the model is trained on the poisoned IRIS dataset with $\Delta L=10\%$. Similarly, for the same dataset, FDR=0.02 at $\Delta L=15\%$ and FDR=0.05 at $\Delta L=20\%$ across all attacks. Similar stability is visible for all adversarially trained models with FORT, as shown in Tables \ref{tab:dt_fdr} and \ref{tab:mlp_fdr}, highlighting the effectiveness of FORT. 
\begin{table}[!ht]
\tiny
    \centering
    \captionsetup[subfloat]{width=0.5\linewidth, justification=centering}
    \caption{Effectiveness of FORT on the FDR of \textbf{RF Model} after poisoning}
    \label{tab:rf_fdr}
    \begin{tabular}{c c c c c c c c}
\toprule
\multirow{2}{*}{Attack} & \multirow{2}{*}{Dataset} & & & \multicolumn{3}{c}{FDR}\\
\cline{3-8}
& & $\Delta L=10\%$ & FORT & $\Delta L=15\%$ & FORT & $\Delta L=20\%$ & FORT \\
\midrule
\multirow{3}{*}{OOP} & IRIS & 0.05 & 0.06 & 0.1 & 0.0001 & 0.19 & 0.13 \\
& MNIST & 0.02 & 0.01 & 0.16 & 0.14 & 0.21 & 0.16 \\
& USPS & 0.09 & 0.04 & 0.15 & 0.08 & 0.2 & 0.09 \\\hline
\multirow{3}{*}{SubP} & IRIS & 0.08 & 0.06 & 0.1 & 0.0001 & 0.21 & 0.07 \\
& MNIST & 0.02 & 0.01 & 0.16 & 0.14 & 0.2 & 0.16 \\
& USPS & 0.1 & 0.04 & 0.16 & 0.08 & 0.2 & 0.09 \\\hline
\multirow{3}{*}{RLP} & IRIS & 0.08 & 0.06 & 0.09 & 0.01 & 0.27 & 0.07 \\
& MNIST & 0.02 & 0.01 & 0.21 & 0.14 & 0.27 & 0.17 \\
& USPS & 0.12 & 0.04 & 0.21 & 0.08 & 0.26 & 0.09 \\
\bottomrule 
    \end{tabular}
    \vspace{0.8cm}
\end{table}

\begin{table}[!ht]
\tiny
    \centering
    \captionsetup[subfloat]{width=0.5\linewidth, justification=centering}
    \caption{Effectiveness of FORT on the FDR of \textbf{DT Model} after poisoning}
    \label{tab:dt_fdr}
    \begin{tabular}{c c c c c c c c}
\toprule
\multirow{2}{*}{Attack} & \multirow{2}{*}{Dataset} & & & \multicolumn{3}{c}{FDR}\\
\cline{3-8}
& & $\Delta L=10\%$ & FORT & $\Delta L=15\%$ & FORT & $\Delta L=20\%$ & FORT \\
\midrule
\multirow{3}{*}{OOP} & IRIS & 0.03 & 0.02 & 0.1 & 0.03 & 0.19 & 0.07 \\
& MNIST & 0.15 & 0.14 & 0.19 & 0.14 & 0.26 & 0.23 \\
& USPS & 0.15 & 0.11 & 0.21 & 0.19 & 0.27 & 0.2 \\\hline
\multirow{3}{*}{SubP} & IRIS & 0.07 & 0.05 & 0.15 & 0.05 & 0.13 & 0.15 \\
& MNIST & 0.14 & 0.14 & 0.19 & 0.15 & 0.26 & 0.23 \\
& USPS & 0.14 & 0.12 & 0.2 & 0.19 & 0.26 & 0.19 \\\hline
\multirow{3}{*}{RLP} & IRIS & 0.15 & 0.03 & 0.12 & 0.05 & 0.23 & 0.11 \\
& MNIST & 0.19 & 0.14 & 0.25 & 0.15 & 0.33 & 0.24 \\
& USPS & 0.19 & 0.12 & 0.26 & 0.19 & 0.34 & 0.22 \\
\bottomrule 
    \end{tabular}
    \vspace{0.8cm}
\end{table}

\begin{table}[!ht]
\tiny
    \centering
    \captionsetup[subfloat]{width=0.5\linewidth, justification=centering}
    \caption{Effectiveness of FORT on the FDR of \textbf{GNB Model} after poisoning}
    \label{tab:gnb_fdr}
    \begin{tabular}{c c c c c c c c}
\toprule
\multirow{2}{*}{Attack} & \multirow{2}{*}{Dataset} & & & \multicolumn{3}{c}{FDR}\\
\cline{3-8}
& & $\Delta L=10\%$ & FORT & $\Delta L=15\%$ & FORT & $\Delta L=20\%$ & FORT \\
\midrule
\multirow{3}{*}{OOP} & IRIS & 0.06 & 0.04 & 0.13 & 0.08 & 0.1 & 0.03 \\
& MNIST & 0.3 & 0.29 & 0.31 & 0.29 & 0.31 & 0.29 \\
& USPS & 0.2 & 0.2 & 0.2 & 0.19 & 0.22 & 0.2 \\\hline
\multirow{3}{*}{SubP} & IRIS & 0.08 & 0.04 & 0.05 & 0.05 & 0.13 & 0.03 \\
& MNIST & 0.29 & 0.29 & 0.32 & 0.28 & 0.3 & 0.28 \\
& USPS & 0.2 & 0.19 & 0.2 & 0.2 & 0.23 & 0.19 \\\hline
\multirow{3}{*}{RLP} & IRIS & 0.06 & 0.04 & 0.11 & 0.05 & 0.12 & 0.03 \\
& MNIST & 0.3 & 0.3 & 0.33 & 0.28 & 0.34 & 0.28 \\
& USPS & 0.21 & 0.19 & 0.22 & 0.19 & 0.24 & 0.2 \\
\bottomrule 
    \end{tabular}
\end{table}

\begin{table}[!ht]
\tiny
    \centering
    \captionsetup[subfloat]{width=0.5\linewidth, justification=centering}
    \caption{Effectiveness of FORT on the FDR of \textbf{MLP Model} after poisoning}
    \label{tab:mlp_fdr}
    \begin{tabular}{c c c c c c c c}
\toprule
\multirow{2}{*}{Attack} & \multirow{2}{*}{Dataset} & & & \multicolumn{3}{c}{FDR}\\
\cline{3-8}
& & $\Delta L=10\%$ & FORT & $\Delta L=15\%$ & FORT & $\Delta L=20\%$ & FORT \\
\midrule
\multirow{3}{*}{OOP} & IRIS & 0.07 & 0.02 & 0.04 & 0.02 & 0.15 & 0.05 \\
& MNIST & 0.06 & 0.01 & 0.06 & 0.02 & 0.08 & 0.02 \\
& USPS & 0.1 & 0.01 & 0.14 & 0.02 & 0.18 & 0.01 \\\hline
\multirow{3}{*}{SubP} & IRIS & 0.03 & 0.03 & 0.05 & 0.02 & 0.2 & 0.07 \\
& MNIST & 0.06 & 0.01 & 0.08 & 0.02 & 0.08 & 0.02 \\
& USPS & 0.1 & 0.02 & 0.13 & 0.01 & 0.16 & 0.01 \\\hline
\multirow{3}{*}{RLP} & IRIS & 0.03 & 0.0001 & 0.07 & 0.04 & 0.37 & 0.05 \\
& MNIST & 0.07 & 0.02 & 0.09 & 0.02 & 0.1 & 0.02 \\
& USPS & 0.1 & 0.01 & 0.13 & 0.01 & 0.16 & 0.02 \\
\bottomrule 
    \end{tabular}
\end{table}
\subsubsection{\textbf{Impact of Increasing Poisoning Rate}} \label{poisoning_rate} SecureLearn maintains effectiveness across all evaluated attacks, independent of increasing poisoning levels. We extended our analysis to understand the relationship between the impact of increasing poisoning levels and the effectiveness of SecureLearn set between $10\% < \Delta L < 20\%$. SecureLearn achieves a minimum sanitized accuracy of 90\% for all models developed with four selected algorithms, highlighting no significant trade-off between model accuracy and adversarial robustness. The results are shown in Fig. \ref{fig:data poisoning with oop attack} to Fig. \ref{fig:data poisoning with rlp attack}. Data poisoning, however, impacts the recall and F1-score differently for each model. The results are given in Table \ref{tab:recall_f1_score}. For RF models, SecureLearn stabilizes these models with a minimum recall of 84.19\% and F1-score of 81.54\% at 20\% OOP poisoning. For DT models, the minimum recall is 78.20\% and the F1-score is 75.80\%. However, it is observed that SecureLearn does not sufficiently stabilizes GNB model trained with the MNIST dataset, as recall remains approximately 57\% and the F1-score to 56\% across poisoning levels. In contrast, SecureLearn is highly effective in securing MLP models, achieving a minimum recall and F1-score of 97\%, which demonstrates its potential to enhance the security of DL models. Overall, these results indicated that SecureLearn effectively mitigates the impact of data poisoning across datasets, even as poisoning levels increase. 

\section{Discussion and Limitations}
\begin{itemize}
\item \textbf{Effects Of Each Component In SecureLearn} We propose SecureLearn as a two-layer defense to mitigate data poisoning attacks and improve the resilience of both traditional ML models and neural networks. SecureLearn proposes an improvised data sanitization along with a generic formulation of adversarial training, considering a common characteristic of the feature importance score. SecureLearn is analyzed and compared with two existing solutions and three data poisoning attacks at three poisoning levels $10\% < \Delta L < 20\%$. The results showed that SecureLearn outperformed others in improving both the security and adversarial robustness of ML against various data poisoning attacks. \\
SecureLearn effectively enhanced the resilience of multiclass ML across RF, DT, GNB and MLP, confirming its generalization beyond algorithm-specific defenses. For all evaluated models, SecureLearn consistently maintains a minimum $90\%$ accuracy and at least 75\% recall and F1-score. SecureLearn successfully reduced the FDR to at least 0.06 against three distinct poisoning attacks. For MLP, SecureLearn achieved a minimum of 97\% recall and F1-score against all selected data poisoning attacks. Furthermore, the adversarial robustness of models is improved with an average accuracy trade-off of $<3\%$. \\
Although various solutions \cite{chen2021pois, baker2024poison, tao2021better} are provided in the literature, none have proposed a two-layer approach. Also, existing adversarial training mechanisms, for example \cite{tao2021better} are limited to gradient-oriented models, which work for neural networks and DL models but are ineffective in proactively securing traditional models against data poisoning attacks. We have taken into account the feature importance of the model and proposed FORT to enhance adversarial robustness of ML. The feature importance score informs the decision criteria of the model and helps generalize the model. By adding a small fraction of perturbation into the features with high importance, the model is taught to distinguish benign and poisoned data points. In this way, the resilience of the model is improved. 
\item \textbf{Limitations} We have experimented SecureLearn to mitigate data poisoning attacks in classification algorithms, which can be further extended to regression algorithms. In this way, we understand the effectiveness and behavior of realigning classifications in multiclass models. Furthermore, implementing SecureLearn in complex deep learning models allows us to understand its efficacy in deep networks, which is out of the scope of this study.  
\end{itemize}
\section{Conclusion} This paper presented SecureLearn, a new attack-agnostic method to defend multiclass ML models from data poisoning attacks. SecureLearn defends against black-box poisoning without prior knowledge of the model and does not require any additional dataset. SecureLearn provides robustness to the model in a two-layer approach, first by sanitizing the training dataset with an improved method and second by enhancing the adversarial robustness with FORT adversarial training. We provided a new approach of adversarial training by developing perturbations with feature importance score rather than gradient learning. This new approach makes adversarial training adaptable to all types of ML and DL algorithms. SecureLearn is applied to four ML algorithms poisoned with three data poisoning attacks, providing promising results. Our results highlight its efficacy against all types of data poisoning attacks, proving it to be an attack-agnostic solution. We also highlight its better performance in most cases compared with existing defenses. Our work improves the understanding of multiclass poisoning and provides an enhanced mitigation to make the training pipelines of ML secure and trustworthy. \\
In the future, we will expand our research and examine SecureLearn in DL and complex ML models, which are used in many digital applications. We also enhanced SecureLearn to improve the security against inference-time poisoning and understand its efficacy in generative AI models and reinforcement learning. 
\bibliographystyle{ieeetr}
\bibliography{References}
\begin{IEEEbiography}[{\includegraphics[width=1in,height=1.25in,clip,keepaspectratio]{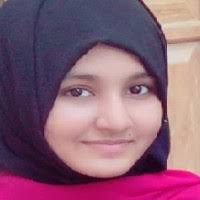}}]{Anum Paracha}
is a PhD student at the School of Computing and Digital Technology, Birmingham City University, UK. Her research interests are to investigate use of advanced machine learning techniques to mitigate emerging cybersecurity research challenges.\end{IEEEbiography}
\begin{IEEEbiography}
[{\includegraphics[width=1in,height=1.25in,clip,keepaspectratio]{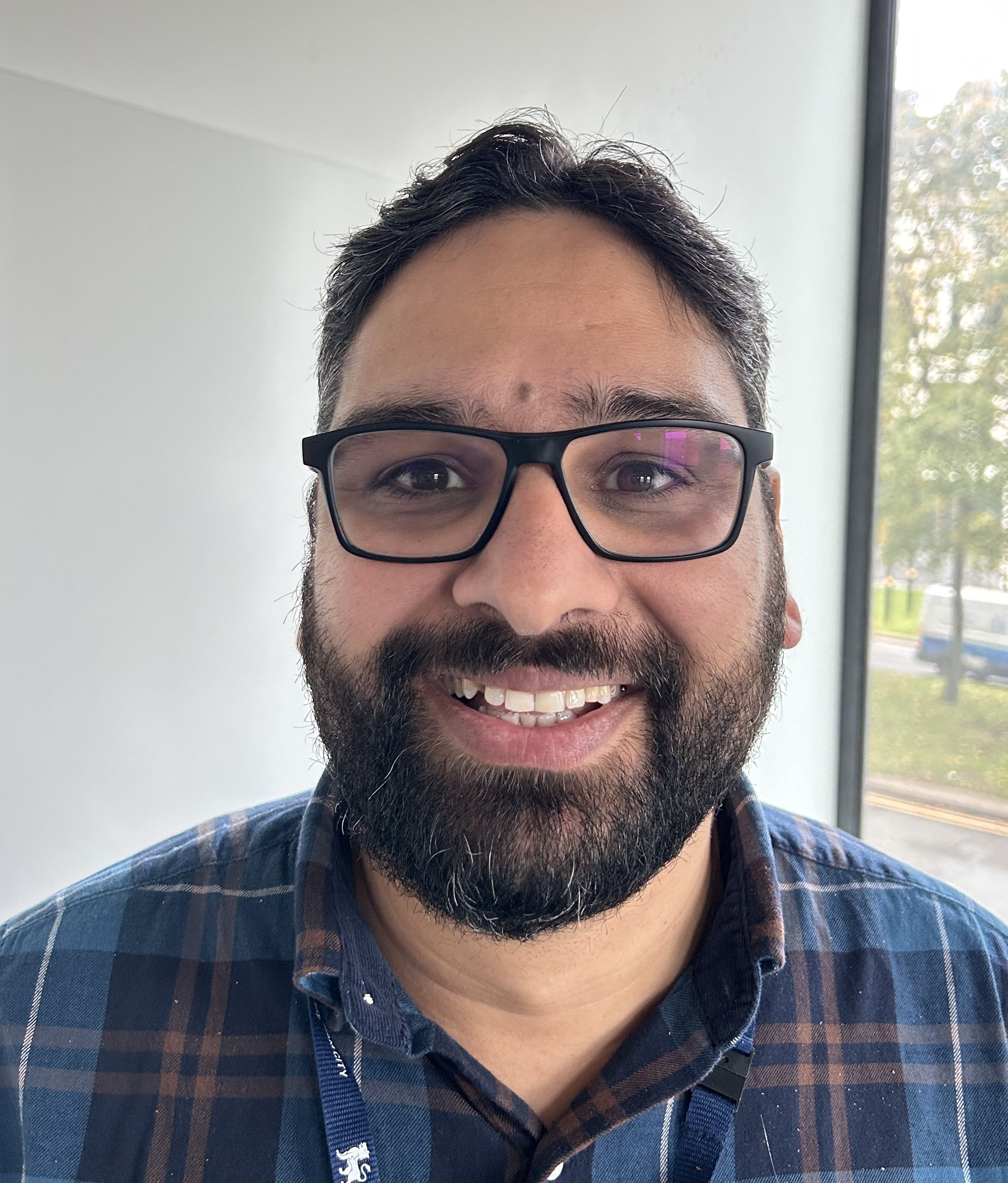}}]{Junaid Arshad}
is a Professor in Cyber Security and has extensive research experience and expertise in investigating and addressing cybersecurity challenges for diverse computing paradigms. Junaid has strong experience of developing bespoke digital solutions to meet industry needs. He has extensive experience of applying machine learning and AI algorithms to develop bespoke models to address specific requirements. He is also actively involved in R\&D for secure and trustworthy AI, focusing on practical adversarial attempts on such systems especially as a consequence of cutting-edge applications of generative AI.\end{IEEEbiography}
\begin{IEEEbiography}
[{\includegraphics[width=1in,height=1.25in,clip,keepaspectratio]{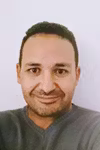}}]{Dr Mohamed Ben Farah} 
is a Senior Lecturer in Cyber Security at Birmingham City University. Mohamed has published over 30 journal and conference papers and has organized conferences and workshops in Cyber Security, Cryptography and Artificial Intelligence. He is a reviewer for world-leading academic conferences and journals and is the Outreach Lead of the Blockchain Group for IEEE UK and Ireland.\end{IEEEbiography}
\begin{IEEEbiography}
[{\includegraphics[width=1in,height=1.25in,clip,keepaspectratio]{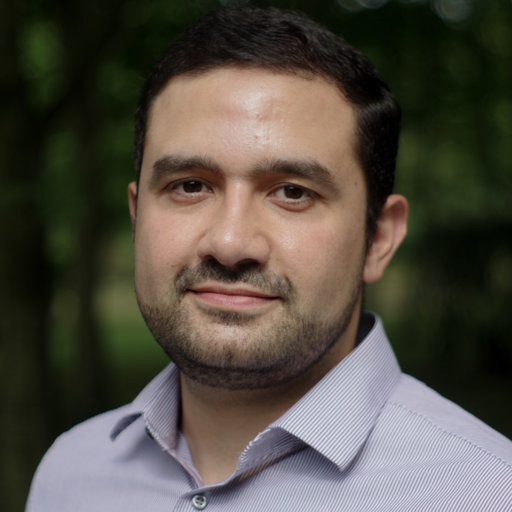}}]{Dr Khalid Ismail}
is a Senior Lecturer in Computer Science at Birmingham City University. His primary research interests lie in the fields of Artificial Intellegence, computer vision, advanced machine learning, image processing, and deep learning, particularly when applied to complex real-world challenges. Currently he is supervising many AI based intelligent projects development and also been an active part of industry based collaborative projects.\end{IEEEbiography}
\end{document}